\documentclass[sigconf,screen]{acmart}
\usepackage[linesnumbered,ruled,vlined]{algorithm2e}
\usepackage{enumitem}
\usepackage[]{collab}

\collabAuthor{zb}{teal}{Zhuangbin Chen}
\collabAuthor{yx}{red}{Yuxin Su}
\collabAuthor{hy}{blue}{hongyu}
\AtBeginDocument{%
  \providecommand\BibTeX{{%
    \normalfont B\kern-0.5em{\scshape i\kern-0.25em b}\kern-0.8em\TeX}}}

\copyrightyear{2022}
\acmYear{2022}
\setcopyright{acmcopyright}
\acmConference[ICSE '22]{44th International Conference on Software Engineering}{May 21--29, 2022}{Pittsburgh, PA, USA}
\acmBooktitle{44th International Conference on Software Engineering (ICSE '22), May 21--29, 2022, Pittsburgh, PA, USA}
\acmPrice{15.00}
\acmDOI{10.1145/3510003.3510085}
\acmISBN{978-1-4503-9221-1/22/05}

\begin{document}

% \title{Adaptive Time Series Anomaly Detection for Online Services via System Failure Sketching}
\title{Adaptive Performance Anomaly Detection for Online Service Systems via Pattern Sketching}

\author{Zhuangbin Chen}
\affiliation{%
  \institution{The Chinese University of Hong Kong}
  \city{Hong Kong}
  \country{China}}

\author{Jinyang Liu}
\affiliation{%
  \institution{The Chinese University of Hong Kong}
  \city{Hong Kong}
  \country{China}}

\author{Yuxin Su}
\affiliation{%
  \institution{School of Software Engineering\\ Sun Yat-sen University}
  \city{Zhuhai}
  \country{China}}
\authornote{Corresponding author (suyx35@mail.sysu.edu.cn).}

\author{Hongyu Zhang}
\affiliation{%
  \institution{The University of Newcastle}
  \city{NSW}
  \country{Australia}}

\author{Xiao Ling}
\author{Yongqiang Yang}
\affiliation{%
  \institution{Huawei Cloud BU}
  \city{Beijing}
  \country{China}}

\author{Michael R. Lyu}
\affiliation{%
  \institution{The Chinese University of Hong Kong}
  \city{Hong Kong}
  \country{China}}

% \affiliation{%

% \affiliation{%


\begin{abstract}

% Recent years have witnessed an increasing prevalence of IT services directly supplied by cloud vendors.
To ensure the performance of online service systems, their status is closely monitored with various software and system metrics. Performance anomalies represent the performance degradation issues (e.g., slow response) of the service systems. When performing anomaly detection over the metrics, existing methods often lack the merit of interpretability, which is vital for engineers and analysts to take remediation actions. Moreover, they are unable to effectively accommodate the ever-changing services in an online fashion. To address these limitations, in this paper, we propose ADSketch, an interpretable and adaptive performance anomaly detection approach based on pattern sketching. ADSketch achieves interpretability by identifying groups of anomalous metric patterns, which represent particular types of performance issues. The underlying issues can then be immediately recognized if similar patterns emerge again. In addition, an adaptive learning algorithm is designed to embrace unprecedented patterns induced by service updates or user behavior changes. The proposed approach is evaluated with public data as well as industrial data collected from a representative online service system in Huawei Cloud. The experimental results show that ADSketch outperforms state-of-the-art approaches by a significant margin, and demonstrate the effectiveness of the online algorithm in new pattern discovery. Furthermore, our approach has been successfully deployed in industrial practice.
%\hy{adaptive is just a part of the proposed approach, maybe no need to say it in the paper title} \zb{yes, but I think putting it in the title can make this merit more direct to the audience and make the paper more interesting. How about we just keep it?}
\end{abstract}

\begin{CCSXML}
<ccs2012>
   <concept>
       <concept_id>10010520.10010521.10010537.10003100</concept_id>
       <concept_desc>Computer systems organization~Cloud computing</concept_desc>
       <concept_significance>500</concept_significance>
       </concept>
   <concept>
       <concept_id>10010520.10010575.10010577</concept_id>
       <concept_desc>Computer systems organization~Reliability</concept_desc>
       <concept_significance>500</concept_significance>
       </concept>
   <concept>
       <concept_id>10010520.10010575.10010579</concept_id>
       <concept_desc>Computer systems organization~Maintainability and maintenance</concept_desc>
       <concept_significance>500</concept_significance>
       </concept>
 </ccs2012>
\end{CCSXML}

\ccsdesc[500]{Computer systems organization~Cloud computing}
\ccsdesc[500]{Computer systems organization~Reliability}
\ccsdesc[500]{Computer systems organization~Maintainability and maintenance}

\keywords{Cloud computing, performance anomaly detection, online learning}


\maketitle

%% Sections
\section{Introduction}
\label{sec:introduction}

% To ensure the continuity of cloud services, KPI anomaly should be captured effectively, which often serve as (early) signals for critical failures. However, accuracy alone is far from satisfaction, as it would be labor-intensive to manually investigate the problematic KPIs for failure understanding. ADSketch facilitates this process by providing prompt anomaly alert with a certain interpretation.

% \zb{expand the first paragraph of Methodology}

% \zb{Title changed as failure profiling is a very old term and we may not align with it} \hy{via System Failure Patterns? Anomaly Patterns?}
% zb: how about we use failure sketching, we have found some paper using this term

With the emergence of cloud computing, many traditional software systems have been migrated to cloud computing platforms as online services. Similar to conventional shrink-wrapped software, the performance of online service systems is an important quality attribute. %Different from conventional shrink-wrapped software, 
As online services need to serve millions of customers worldwide, a short period of performance degradation could lead to economic loss and user dissatisfaction. 
%A short period of service downtime could lead to economic loss and user dissatisfaction. 
Therefore, proactive and even adaptive system troubleshooting has become the core competence of online service providers. %As data-driven by nature, AIOps (Artificial Intelligence for IT operations)~\cite{dang2019aiops,aiops} is deemed as a promising technique to address the challenges of IT operations.
Enterprises that have promoted the automation of system troubleshooting have already received real gains in reliability, efficiency, and agility~\cite{lou2013software,chen2020towards,dang2019aiops}.

In industrial scenarios, online service systems are closely monitored with various metrics (e.g., the CPU usage of an application, service response delay) on a 24$\times$7 basis. This is because the monitoring metrics often serve as the most direct and fine-grained signals that flag the occurrence of service performance issues. In addition, they provide informative clues for engineers to pinpoint the root causes. However, due to the large scale and complexity of online service systems, the number of metrics is overwhelming the existing troubleshooting systems~\cite{chen2020towards}. Automated anomaly detection over the metrics, which aims to discover the unexpected or rare behaviors of the metric time series, is therefore an important means to ensure the reliability and availability of service systems.
% \hy{I think the proposed approach works for general large-scale software systems? Online service systems could be one type of such software systems and the performance problem is more important...} \zb{Yes, indeed the proposed approach works for general software systems. Using the context of online services is because we are working with Huawei cloud and basically all datasets I can find are about services... I will mention that ADSketch is actually very general.}

Although many efforts, e.g.,~\cite{zhao2021predicting,su2019robust,he2018identifying}, have been devoted to performance anomaly detection, most of the existing work does not possess the merit of interpretability. Specifically, at each timestamp, they calculate a probability indicating the likelihood of performance anomalies. A threshold is then chosen to convert the probability into a binary label – normal vs. anomaly. However, in reality, a simple recommendation of the suspicious anomalies might not be of much interest to engineers. 
This is because they need to manually investigate the problematic metrics (recommended by the model) for fault localization. 
For large-scale online services, this process is like finding a needle in a haystack. The problem is compounded by the fact that false alerts are not rare. Moreover, many state-of-the-art methods train models with historical metric data in an offline setting. As online services continuously undergo feature upgrades and system renewal, the patterns of metrics may evolve accordingly, i.e., concept drift~\cite{gama2014survey,han2020toward}. Without adaptability, these models are unable to accommodate the ever-changing services and user behaviors. %Our experiments have revealed that with frozen data and parameters (including the threshold), their performance can drop significantly in online scenario where metric time series arrive in streams.

In this paper, we propose ADSketch, a performance anomaly detection approach for online service systems based on \textit{pattern sketching}, which is interpretable and adaptive. %Moreover, ADSketch requires no threshold to make the final anomaly decision for a data point, which improves its simplicity and determinacy. 
%Different from many existing models (e.g., deep neural networks) that implicitly simulate the normal behaviors of a KPI, 
%\hy{can introduce failure sketching first} \zb
The main idea is to identify discriminative subsequences from metric time series that can represent classes of service performance issues. This is similar to the problem of shapelet discovery in time series data~\cite{rakthanmanon2013fast,yeh2016matrix}. Particularly, for multiple subsequences that describe the same type of performance issue, we take the average of them and regard the result as a \textit{metric pattern} for the issue.
% extract anomalous metric patterns (i.e., time series segments corresponding to the misbehaving moments of metrics) that can represent service performance issues.
For example, services may be experiencing performance degradation when we observe a level shift down on Service Throughput or a level shift up on CPU Utilization. The advantages of such metric patterns are twofold. First, the normality of the incoming metric subsequences can be quickly determined through a comparison with the metric patterns.
% Subsequences that carry a normal pattern will be regarded as anomaly-free.
Second, by associating the patterns with typical anomaly symptoms, we can immediately understand the ongoing performance issues when the metric subsequences exhibit known patterns. This is similar to failure/issue profiling~\cite{podgurski2003automated,lin2016log,ma2020diagnosing}.
% sufficiently
% We explicitly consider the \textit{context} of the KPI by extracting its normal and and archive them in a storage-friendly way.
% Moreover, as anomalous patterns are often triggered by particular types of failures, they can be used as an indicator for the corresponding failures, i.e., failure sketching.
% \hy{the term failure sketching should be explained}
In this way, ADSketch provides a novel mechanism to characterize service performance issues with metric time series. Previous work on failure/issue profiling often requires handcrafted features, which suffers from limited generalization. For example, Brandon et al.~\cite{brandon2020graph} manually defined a set of features collected from metrics, logs, and anomalies to characterize failures. Pattern sketching with metrics enjoys the advantages of automation and accuracy. Moreover, ADSketch is able to adaptively embrace new anomalous patterns when detecting anomalies on the fly. Experimental results demonstrate the superiority of our design over the existing state-of-the-art time series anomaly detectors on both public and industrial data. In particular, we have achieved an average F1 score of over 0.8 in production systems.
%\hy{briefly introduce the experiments and results}

To sum up, this work makes the following major contributions:

% [noitemsep,topsep=0pt]
\begin{itemize}
    % \item We propose the concept of pattern sketching, which offers a way to characterize service performance issues with the time series of monitoring metrics. Pattern sketching can facilitate the process of performance issue understanding and mitigation for engineers. Particularly, we also design efficient algorithms for metric pattern extraction and evolution in online scenarios.\hy{to me, it looks like Pattern Mining/Discovery..not sure if concept of "pattern sketching" is a contribution (and if it is needed)}
    
    \item We propose ADSketch, an interpretable and adaptive approach for service performance anomaly detection. ADSketch offers a way to characterize service performance issues with monitoring metrics. Different from the existing work, ADSketch is able to provide explanations (e.g., the type of the underlying performance issues) for its prediction results and accept new patterns on the fly. The implementation of ADSketch and datasets are publicly available on GitHub\footnote{\url{https://github.com/OpsPAI/ADSketch}}.
    % \textcolor{red}{Moreover, ADSketch requires no threshold to make anomaly decisions, which improves its simplicity and determinacy.}
    
    \item We conduct experiments with public data as well as industrial service metric data collected from Huawei Cloud. The experimental results demonstrate the effectiveness of ADSketch in terms of both anomaly detection and adaptive metric pattern learning. Furthermore, our framework has been successfully incorporated into the service performance monitoring system of Huawei Cloud. Our industrial practice confirms its practical usefulness.
\end{itemize}

%The remainder of this paper is organized as follows. Section~\ref{sec:background} introduces the background and problem statement of this paper. Section~\ref{sec:methodology} elaborates on the algorithms for adaptive performance anomaly detection. Section~\ref{sec:exp} presents experiments and experimental results. Section~\ref{sec:related_work} discusses some related work. Section~\ref{sec:discussion} shares our success story and some case studies. Finally, Section~\ref{sec:conclusion} concludes this work.

\section{Background \& Problem Statement}
\label{sec:background}

\subsection{Performance Anomaly Patterns in Online Service Systems}
\label{sec:failure_pattern}

% To ensure the continuity of online services, cloud systems often possess an abundance of redundant components, providing the ability of fault tolerance. Consequently, the majority of performance anomalies and availability breakdowns in cloud environments tend to be caused by subtle underlying faults, i.e., gray failures instead of fail-stop failures~\cite{huang2017gray}. 

In online service systems, a large number of metrics are configured to monitor various aspects of both logical resources (e.g., a virtual machine) and physical resources (e.g., a computing server).
% \zb{the characteristics of software (modern/complicated software/services): self-healing (e.g., load balancing), can be found by our method (in the form of anomalous patterns which disappear quickly)}
Cloud systems often possess an abundance of redundant components, providing the ability of fault tolerance and self-healing (e.g., load balancing, availability zones). Consequently, the majority of service breakdowns tend to manifest themselves as performance anomalies first instead of fail-stop failures~\cite{huang2017gray,lou2020understanding}. We observe when performance anomalies of similar types happen, their impacts tend to trigger similar reactions/symptoms on the metric time series, which we refer to as metric patterns. For example, a level shift up on Interface Throughput may indicate slow service response, which could be caused by a load balancing failure; a level shift down on it may suggest service unavailability, and the culprit could be performance bugs (e.g., memory leak bugs). Similar observations have been made in \cite{ma2020diagnosing,farshchi2015experience}. The rationale behind such a phenomenon is twofold. First, the design of the metrics is sophisticated and fine-grained, each of which is dedicated to monitoring a specific problem, e.g., request timeout, high API error rate. Second, cloud systems widely employ the microservices architecture, where cloud applications employ lightweight container deployment, e.g., cloud-native applications, serverless computing. With this architecture, each microservice is designated for well-defined and modularized jobs, e.g., user login, location service. Thus, they tend to develop individual and stable patterns, which can manifest through their monitoring metrics.

\begin{figure}[t]
    \centering
    \includegraphics[width=0.9\linewidth]{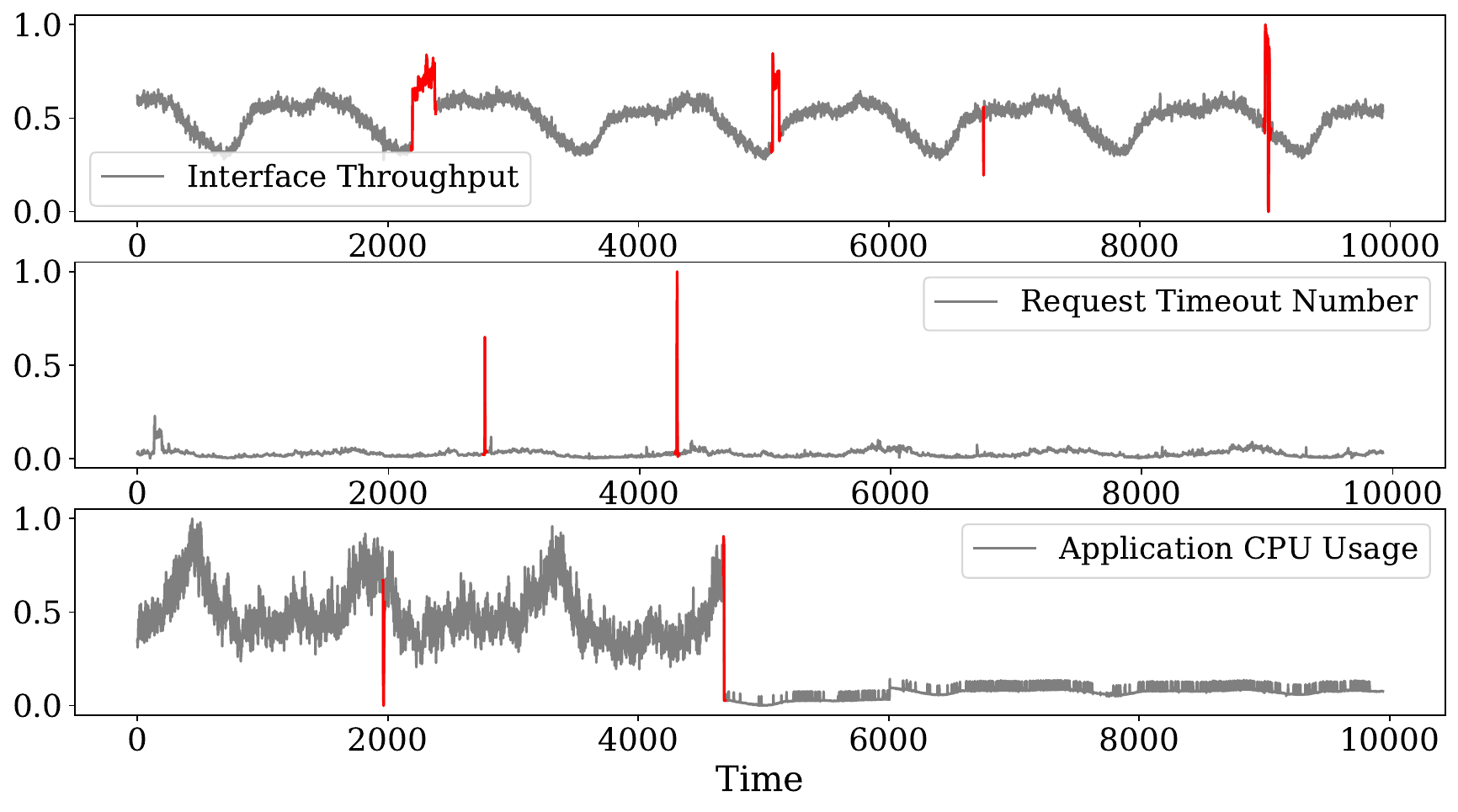}
    \caption{Examples of performance anomaly patterns}
    \vspace{-0.04in}
    \label{fig:anomaly_patterns}
    \vspace{-0.06in}
\end{figure}

% For example, in normal cases, the metric of port in-bound/out-bound traffic rate should float within appropriate ranges. Spike shift is one typical anomalous pattern on this metric time series.

\subsection{Metric Pattern Mining}

% \hy{is this Pattern Sketching or Pattern Mining/Discovery? or Metric Pattern Mining/Discovery?} \zb{I should mention anomaly detection here because it is the main topic of this paper. I will talk more about the AD part.}\hy{maybe Metric Pattern Mining? some people also use the term fingerprint} \zb{Ok. I think it also fits the stuff this para. talks about.}

Metric patterns (i.e., time-series subsequences describing the misbehaving moments of metrics) can be leveraged to sketch the performance issues for anomaly detection. This is essentially profiling the mode of recurrent anomalies. For example, hardware failures often come with a sudden drop in the corresponding metrics, and the value remains zero for some time. If anomalies come into existence, they can be immediately identified by matching the established patterns. Such metric patterns can also facilitate problem mitigation. For example, when low service throughput and high CPU usage are detected, engineers can scale up the microservice (by adding local cores) to increase its capacity. The key challenge is how to automatically discover what anomalous patterns a metric time series has experienced. For each identified pattern, engineers can label the typical performance issues it often associates with. In online scenarios, if a metric encounters any known anomalous patterns, the underlying performance issues can be recommended. Pattern sketching therefore provides a means to accumulate and utilize engineers' knowledge.

In real-world scenarios, the patterns exhibited in metrics are extremely complicated and can have numerous variants in terms of scale, length, and combination. Particularly, we have identified the following challenges for metric pattern discovery, which are illustrated in Fig.~\ref{fig:anomaly_patterns}. Each metric time series records around one week of monitoring data, whose anomalies are shown in red.

\begin{figure}[t]
    \centering
    \includegraphics[width=0.96\linewidth]{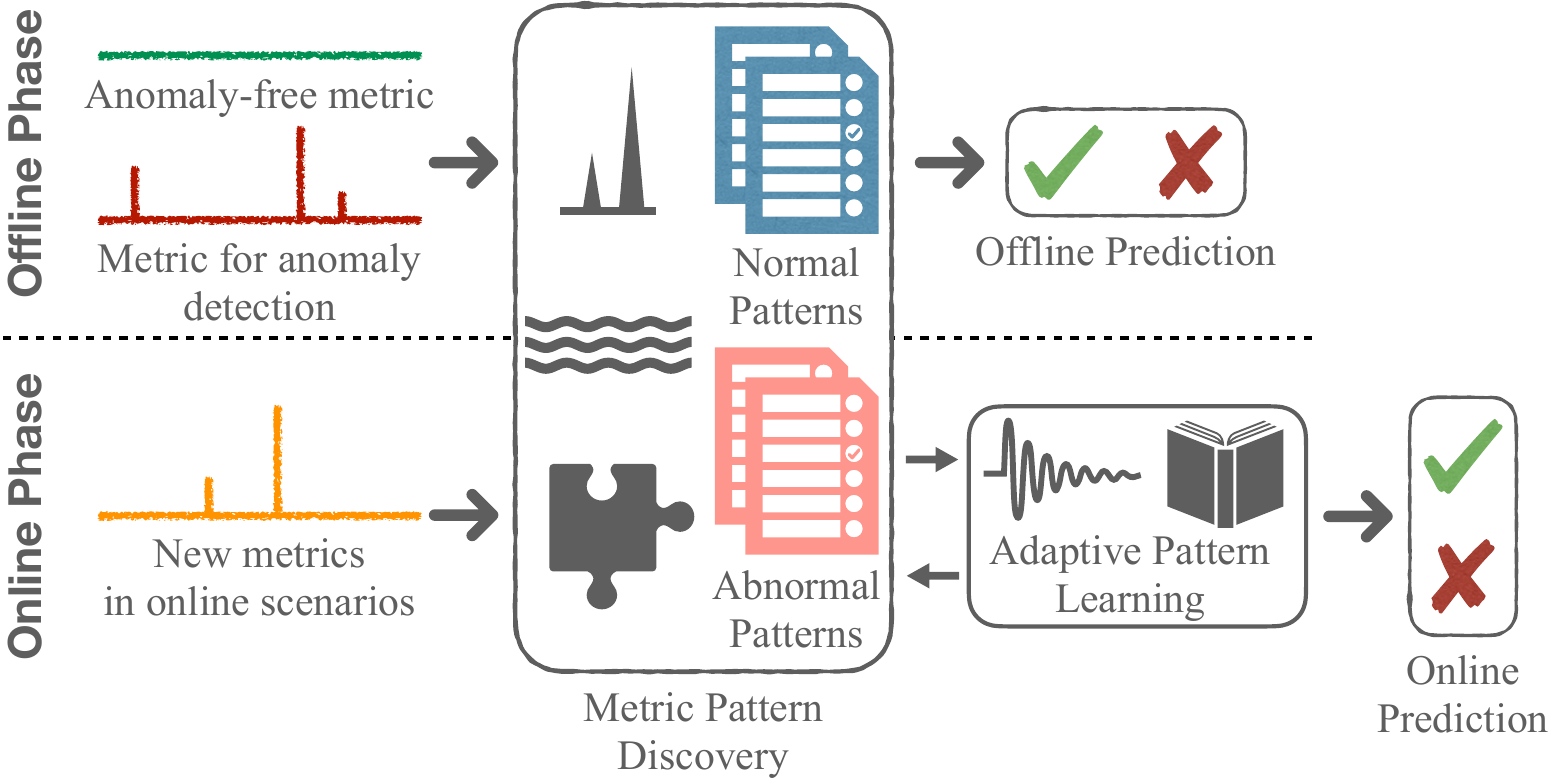}
    \caption{The Overall Framework of ADSketch}
    \label{fig:framework}
    % \vspace{-0.2in}
\end{figure}

% \hy{can describe more about this figure here. For example, the red lines, where are the anomaly patterns, etc}

\textbf{Background noise}. Although a large amount of metric time series is generated, a significant portion of them is trivial, which only records plain system runtime behaviors. Moreover, due to the dynamics of online services, some metrics may experience concept drift~\cite{gama2014survey}. For example, the Application CPU Usage in Fig.~\ref{fig:anomaly_patterns} drops abruptly, which could be caused by a role switch (e.g., from a primary node to a backup node) or user behavior change. How to distinguish anomalous patterns from normal ones is non-trivial.

\textbf{Pattern variety}. A metric curve can possess multiple distinct patterns simultaneously. For example, in Fig.~\ref{fig:anomaly_patterns}, the Interface Throughput has two anomaly patterns, i.e., spike up and spike down. Also, the patterns can have different scales, as indicated by the two spikes in the Request Timeout Number. We need to consider the context of each metric for pattern extraction.

\textbf{Varying anomaly duration}. Different performance issues may vary in duration. The first two anomalies in the Interface Throughput constitute such an example. Particularly, how long an anomaly lasts is also an important factor that engineers rely on to understand a service's health state. When characterizing the issues, such a fact should be properly considered.

% \zb{emphasize the novelty of profiling failures with KPIs (failure profiling is important)}

% In particular, our algorithm provides a concrete method to semantically interpret KPIs.

\subsection{Problem Statement}

The goal of this work is to detect performance anomalies for modern software systems, especially online service systems,
% \hy{see if it can be applied to general software systems}
based on monitoring metrics. To facilitate issue understanding and problem mitigation, we intend to improve the interpretability of the detection results. To this end, we propose to sketch performance issues with metrics based on our observation that similar issues often exhibit alike patterns. By extracting such anomalous metric patterns, we can conduct performance anomaly detection by examining whether the incoming metric subsequences match the known patterns. Moreover, by associating the extracted metric patterns to specific performance issues, we can obtain a quick understanding of the ongoing issues in online scenarios. Additionally, as online services are continuously evolving, unprecedented metric patterns may emerge. Thus, our algorithm should be adaptive to the new patterns. The problem can be formally defined as follows.

The input of a metric time series can be represented as $\mathcal{T}\in \mathbb{R}^l=[t_1, t_2, ..., t_l]$, where $l$ is the number of observations. $t_i^m=[t_i, ..., t_{i+m-1}]$ is a consecutive subsequence of $\mathcal{T}$ starting from $t_i$ with length $m$, where $i\in [0, l-m]$. The objective of performance anomaly detection is to determine whether or not a given $t_i^m$ is anomalous, i.e., whether there are performance issues happening from timestamp $i$ to $i+m-1$. Particularly, we also try to explain the type of performance issues associated with $t_i^m$. The anomalous subsequences will be used to construct abnormal metric patterns, while the benign ones will be regarded as normal patterns. Both the normal and abnormal metric patterns will be updated as the anomaly detection proceeds.

% \textbf{Stringent speed and accuracy requirements}. \zb{for point adjustment, we cannot know 1) the occurrence of failures immediately; 2) how long the impact of failure lasts; 3) }

\section{Methodology}
\label{sec:methodology}

\subsection{Overview}

\begin{table}[t]
\begin{center}
\caption{Summary of Variables}
\label{tab:variable}
\normalsize
 \begin{tabular}{c|l}
  \toprule
  \textbf{Variable} & \textbf{Meaning} \\
  \midrule
  \midrule
  $\mathcal{T}_n$ & An anomaly-free metric time series \\
  $\mathcal{T}_a$ & An input metric time series for anomaly detection \\
  $t$ & A subsequence of metric time series \\
  $m$ & The length of the metric subsequence $t$ \\
  $p$ & The percentile threshold to find deviated subseqs \\
  $\mathcal{P}_n$ & The index set of normal metric patterns \\
  $\mathcal{P}_a$ & The index set of anomalous metric patterns \\
  $\mu_C$ & The vector of cluster mean vectors \\
  $\mathcal{S}_C$ & The vector of cluster sizes \\
  $\mathcal{R}_C$ & The vector of cluster radii \\

  \bottomrule
 \end{tabular} 
\end{center}
\end{table}

\begin{figure*}[t]
    \centering
    \includegraphics[width=1.0\linewidth]{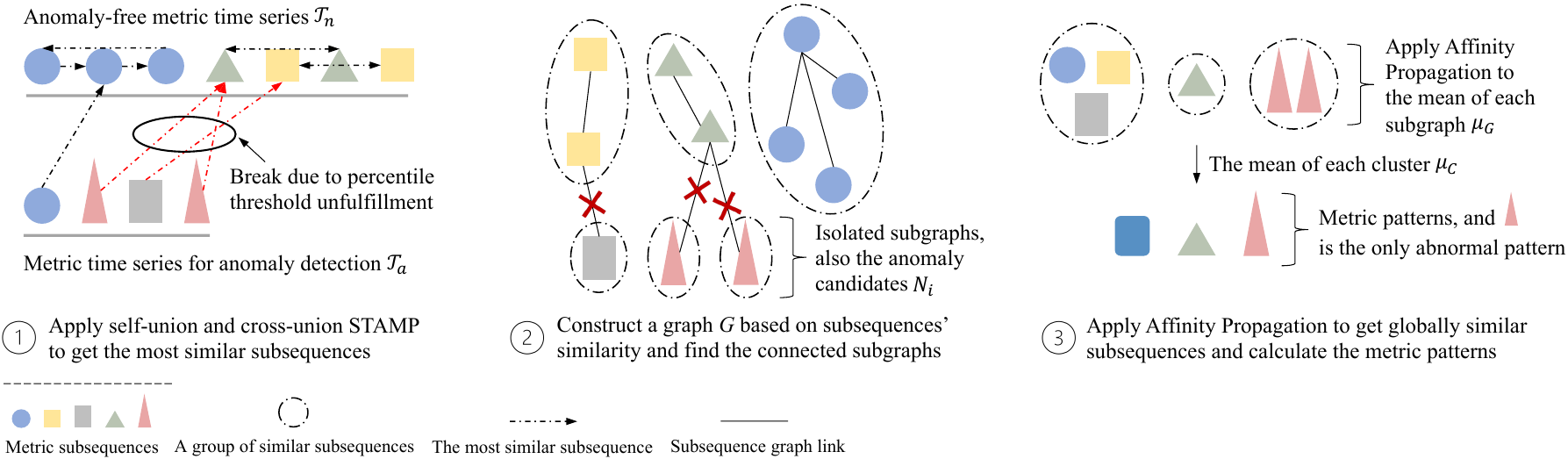}
    % \vspace{-0.02in}
    \caption{The Algorithm of Performance Anomaly Pattern Discovery}%\hy{the notations in the first row are too big, I though they belong to the processing steps too...The figure is a bit too busy...can also add some white space between two steps}\hy{can move the annotations down and put it in a box (or a separation line)?} \zb{I make them like a footnote.}\hy{step 2 "construct a graph", where the graph is? anyway, I suggest to draw a better Figure 2. Also, if space allows, can have 2 figures: one for Overall workflow, one for Algorithm 1.} \zb{The graph is added. I will see if the framework diagram is possible.}}
    \label{fig:adsketch_algorithm1}
    % \vspace{-0.08in}
\end{figure*}

%To ensure the continuity of cloud services,
% To improve the reliability and availability of cloud services, performance anomalies should be detected effectively.
In online service systems, performance anomalies often serve as the (early) signals for critical failures, which should be detected effectively. However, accuracy alone is far from satisfactory, as it will be labor-intensive to manually investigate the problematic metrics for issue understanding. ADSketch facilitates this process by providing prompt anomaly alerts with explanations. 
%a certain interpretation.

% \zb{make clear that ADSketch first conducts anomaly detection, during which normal and abnormal patterns are identified. We can use the identified patterns for offline anomaly detection}
The overall framework of ADSketch is shown in Fig.~\ref{fig:framework}, which consists of two phases, namely, \textit{offline anomaly detection} and \textit{online anomaly detection}. In the offline phase, ADSketch takes as input a pair of metric time series. One metric time series is anomaly-free, which serves as the basis to detect anomalies in the other metric (if any). In this process, a set of metric patterns will be automatically learned. A metric pattern is essentially the mean of a set of similar metric subsequences representing similar service behaviors. The identified metric patterns are divided into two types, i.e., normal and abnormal. The abnormal patterns often characterize some particular types of performance issues, as discussed in Sec.~\ref{sec:failure_pattern}. Thus, by investing manual efforts to link them to the corresponding issues, a clearer picture of the underlying problems can be easily obtained if similar patterns are encountered again. In the online phase, we leverage the metric patterns built in the offline phase to conduct anomaly detection in online scenarios, where metrics arrive in streams. Particularly, in production environments, unprecedented patterns could appear. Thus, we design an adaptive learning algorithm to capture the new patterns continuously.

% \hy{can draw a small overall diagram}

Before formally introducing our algorithms, we have summarized the variables involved in Table~\ref{tab:variable}.

\begin{figure}[t]
    \centering
    \includegraphics[width=0.92\linewidth]{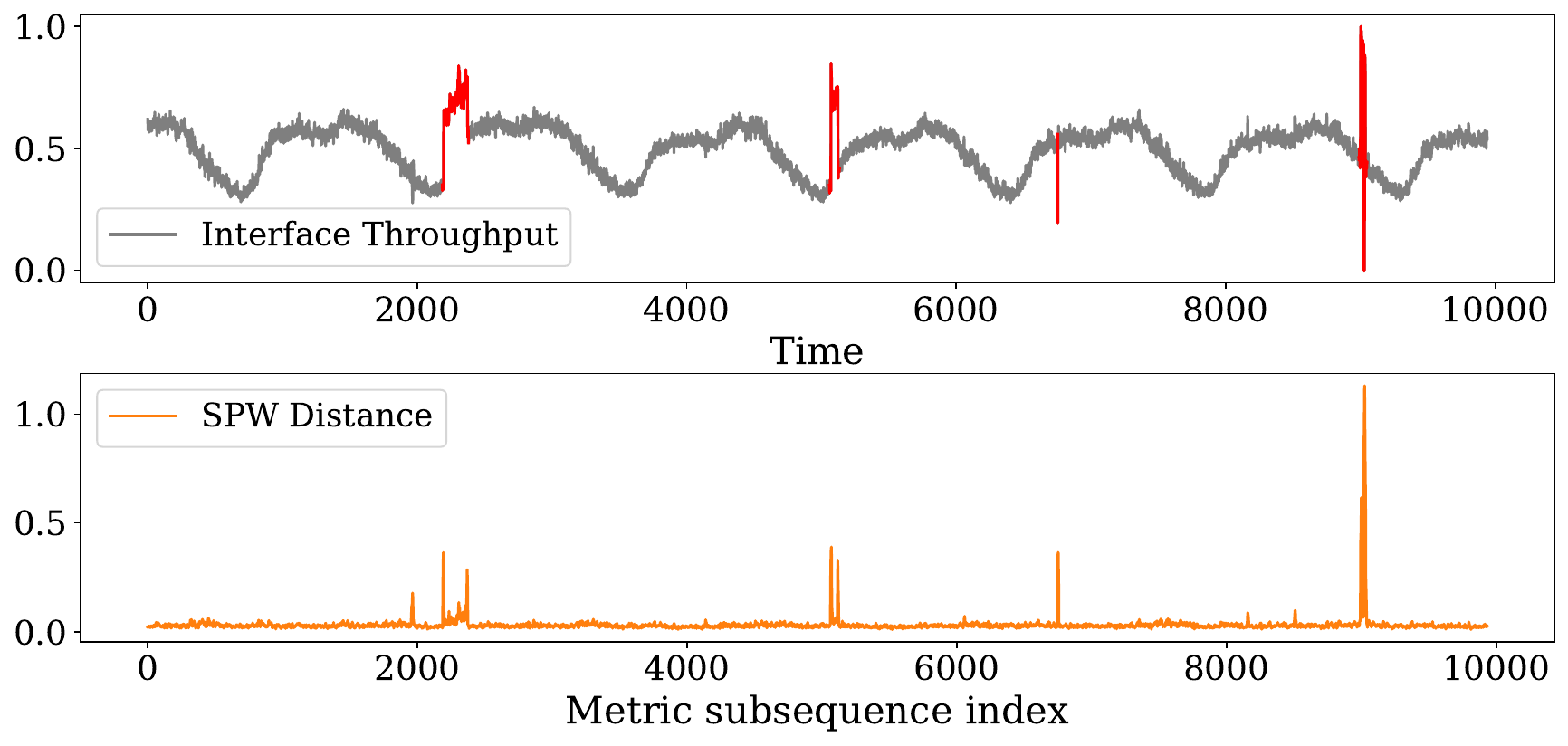}
    \caption{The SPW distance of different metric subsequences}
    \label{fig:mp_example}
    % \vspace{-0.2in}
\end{figure}

\subsection{Offline Anomaly Detection}
\label{sec:offline_anomaly_detection}

\subsubsection{Metric Pattern Discovery}

The idea for discovering the abnormal patterns follows the basic definition of an anomaly: if a metric subsequence deviates significantly from those collected during a service's normal executions, it is likely that the subsequence captures some misbehaving moments of the service. To measure how deviated a metric subsequence is, we calculate its distance to other subsequences and search for the smallest distance score. Intuitively, metric subsequences which have large scores to others tend to be anomalous. The function for distance measure is customizable, and we adopt Euclidean distance in this paper. % It effectively searches for metric segments that share similar shapes. %based on the pairwise distance between all segments.
% The length of a metric segment equals to a pre-defined number $m$. \textcolor{red}{Although the length is fixed, we are able to construct failure patterns longer than $m$, which will be demonstrated in the next section.}

Given a metric time series with $l$ observations, the number of all possible subsequences is $l-m+1$, where $m$ is the length of its subsequences. A naïve solution for calculating the smallest pair-wise distance (which we refer to as \textit{SPW distance} hereafter) would be brute force searching. However, this algorithm owns a quadratic time complexity, which is practically infeasible for large time series. Fortunately, some novel scalable algorithms~\cite{yankov2008disk,yeh2016matrix,zhu2018matrix} have been proposed in the literature to attack such all-pairs-similarity-search problems for time series subsequences. Particularly, Yeh et al.~\cite{yeh2016matrix} proposed STAMP, which has achieved orders of magnitude faster compared to state-of-the-art methods. For exceptionally large datasets, an ultra-fast approximate solution is also provided.
% In STAMP, the smallest pair-wise distance of a subsequence is referred to as matrix-profile distance.
An illustrating example is provided in Fig.~\ref{fig:mp_example}, where we can see the misbehaving metric subsequences have larger SPW distances. In particular, the original STAMP algorithm adopts z-normalization for data preprocessing. However, we found min-max normalization yields more meaningful results in our scenario. For a subsequence $t_i^m$ in a metric time series $\mathcal{T}$, we record the index and distance score of another subsequence having the SPW distance to it. Such index and score of all subsequences, i.e., $t_i^m~(i\in [0, l-m])$, constitute two vectors $\mathcal{I}$ and $\mathcal{S}$. In particular, for $t_i^m$, its closest subsequence can either come from the same time series (i.e., self-union) or another time series (i.e., cross-union). In the first case, a trivial match region around $t_i^m$ will be excluded to avoid self matches~\cite{yeh2016matrix}.
% \hy{maybe can use an illustrating graph to explain Algorithm 1}

The proposed algorithm for metric pattern discovery is presented in Algorithm~\ref{algo:pattern_discovery}, which is illustrated in Fig.~\ref{fig:adsketch_algorithm1}. Algorithm~\ref{algo:pattern_discovery} takes as input two metric time series, i.e., $\mathcal{T}_n$ and $\mathcal{T}_a$ ($\mathcal{T}_n$ is anomaly-free and $\mathcal{T}_a$ may contain anomalies to be detected), and two hyper-parameters, i.e., $m$ and $p$ ($m$ is the length of subsequences and $p$ is the percentile threshold to find the deviated subsequences). As production service systems are mostly running in normal status~\cite{chen2020towards}, the anomaly-free input is easily obtainable (we discuss how we address the violating cases in Sec.~\ref{sec:threats_to_validity}). In line 1 of Algorithm~\ref{algo:pattern_discovery}, we apply STAMP to $\mathcal{T}_n$ with \textit{self-union} (i.e., similar subsequences come from $\mathcal{T}_n$), and obtain the index and score vectors $\mathcal{I}_{nn}$ and $\mathcal{S}_{nn}$. In line 2, we search similar subsequences for $\mathcal{T}_a$ from $\mathcal{T}_n$, i.e., \textit{cross-union}, and get $\mathcal{I}_{na}$ and $\mathcal{S}_{na}$. Intuitively, given the fact that $\mathcal{T}_n$ is anomaly-free, subsequences in $\mathcal{T}_a$ having large SPW distances to their closest peers in $\mathcal{T}_n$ are suspected to be anomalous. Interestingly, we later learn that Mercer et al.~\cite{mercer2021matrix} proposed a similar idea concurrently. We introduce a percentile threshold (i.e., $p$) on $\mathcal{S}_{na}$ to find such deviated subsequences. In particular, $p$ is loosely set to avoid missing anomalies, i.e., false negatives. Such a setting will inevitably produce false positives. We next discuss how we alleviate this issue.

A metric pattern is defined as the mean of a group of similar subsequences, which represents some typical behaviors of the metric time series. To mine similar subsequences, we propose to leverage their similarity connections. Specifically, in line 3, we construct a graph $G$ whose nodes correspond to the subsequences. Two nodes will be linked if any one of them is deemed as the most similar subsequence to the other, as indicated by $\mathcal{I}_{nn}$ and $\mathcal{I}_{na}$. Note such a relationship is not mutual, i.e., $t_i^m$ is the most similar to $t_j^m$ does not necessarily imply the opposite case. We break the edges whose distance score fails to meet the threshold requirement $p$. The above operations are depicted in the first part of Fig.~\ref{fig:adsketch_algorithm1}. Next, we find the connected subgraphs of $G$, each of which is composed of subsequences resembling each other. Particularly, there will be some isolated nodes, i.e., subgraphs with a single node, which are collected at line 4. Such deviated subsequences constitute a set of anomaly candidates, i.e., $N_i$. The second part of Fig.~\ref{fig:adsketch_algorithm1} illustrates this process.

\begin{algorithm}[t]
\caption{Performance Anomaly Pattern Discovery}
\label{algo:pattern_discovery}
\normalsize
\SetAlgoLined
\KwIn{$\mathcal{T}_n$, $\mathcal{T}_a$, $m$, and $p$}
\KwOut{Two disjoint sets of $\mathcal{P}_n$ and $\mathcal{P}_a$}

% $l_n\gets {\rm Length}(\mathcal{T}_n), l_a\gets {\rm Length}(\mathcal{T}_a)$

$\mathcal{I}_{nn}, \mathcal{S}_{nn}\gets {\rm STAMP}(\mathcal{T}_n, \mathcal{T}_n, m)$

$\mathcal{I}_{na}, \mathcal{S}_{na}\gets {\rm STAMP}(\mathcal{T}_n, \mathcal{T}_a, m)$

$G\gets {\rm ConnectedSubgraphs}(\mathcal{I}_{nn}+\mathcal{I}_{na}, \mathcal{S}_{na}, p)$

$N_i\gets {\rm IsolatedNodes}(G)$

$\mu_G\gets {\rm GraphWiseMean}(G)$

$C\gets {\rm AffinityPropagation(\mu_G)} $ 

$\mu_C\gets {\rm ClusterWiseMean}(C)$

$\mathcal{P}_n\gets {\rm EmptyArray}, \mathcal{P}_a\gets {\rm EmptyArray}$

\For{$each~idx~in~1:{\rm Size}(C)$}{
    \tcp{$C[idx]$: all subsequences in the cluster}
    \eIf{$C[idx]\subset N_i$}{
        $\mathcal{P}_a\gets {\rm Append}~\mathcal{P}_a~{\rm with}~idx$
    }{
        $\mathcal{P}_n\gets {\rm Append}~\mathcal{P}_n~{\rm with}~idx$
    }
}
\end{algorithm}

\begin{algorithm}[t]
\caption{Performance Anomaly Detection}
\label{algo:anomaly_detection}
\normalsize
\SetAlgoLined
\KwIn{$t$, $\mathcal{P}_a$, and $\mu_C$}
\KwOut{Anomaly detection result for $t$}

$\mathcal{D}_t\gets {\rm PairWiseDistance}(t, \mu_C)$

$idx\gets {\rm MinIndex}(\mathcal{D}_t)$

\eIf{$idx\in \mathcal{P}_a$}{
    return True
}{
    return False
}
\end{algorithm}

% However, for subsequences (especially the anomalous ones) in $T_a$, the most similar peer may not be truly similar. Therefore, we set a (loosely) similarity threshold ($p$ percentile of $\mathcal{S}_{na}$ in this paper) to filter out dissimilar (i.e., false) connections in the graph, i.e., cut the edges.

Up to this point, we have divided the subsequences of $\mathcal{T}_n$ and $\mathcal{T}_a$ into different parts, each of which is represented as a subgraph. However, each subgraph cannot be directly regarded as a metric pattern because: 1) the graph construction criteria can be too strict (i.e., only the most similar pairs are connected), so some subgraphs might still be similar; 2) the loosely set percentile threshold $p$ may flag some normal subsequences as abnormal (i.e., false positives). To further combine the similar subsequences, we apply the Affinity Propagation algorithm~\cite{frey2007clustering} to cluster the mean vector of each subgroup (line 5-6). We choose this algorithm because of its superior performance and efficiency, and it requires no pre-defined cluster number. As a result, similar normal subgraphs can be merged together, and abnormal subgraphs have a chance to embrace their normal communities. Thus, each cluster will contain all similar subsequences across the two time-series inputs and different clusters represent distinct patterns. The mean of clusters (i.e., $\mu_C$) will form the set of metric patterns (line 7). For each cluster, we check whether or not all its members come from the set of anomaly candidates $N_i$ (line 9-15). If yes, the mean of the cluster will be regarded as an abnormal metric pattern and otherwise normal, indexed by $\mathcal{P}_a$ and $\mathcal{P}_n$, respectively. The third part of Fig.~\ref{fig:adsketch_algorithm1} presents the above operations. Finally, all subsequences in the anomalous clusters will be predicted as an anomaly to be the output of this phase.

% \zb{actually we can conduct anomaly detection right at this point}

\subsubsection{Metric Pattern Interpretability}
\label{sec:metric_pattern_interpretability}
% \hy{is this section necessary?} \zb{Let me try if the paper still has some room for it}

In this section, we expound on how to label the performance issues that each metric pattern represents.
% The anomalous metric patterns often characterize particular types of performance issues. \textcolor{red}{For example, a spike shift down in port traffic rate could stem from the status change of network protocol.}
By allowing metric patterns to have semantics, the understanding and mitigation of service problems can be greatly accelerated. % In the meanwhile, engineers' system troubleshooting experience can be accumulated, constituting a failure pattern database.
Given the fact that the duration of different performance issues may vary, our fixed-length metric patterns may over-represent (i.e., the metric pattern is much larger than the issue's duration) or under-represent (i.e., the metric pattern is only an excerpt of the issue) the corresponding issues. To alleviate the first problem, we select a relatively small $m$, which turns out to be aligned with the goal of better performance. For the second problem, we adopt the following strategy to group clusters which are actually describing a common issue. For each pair of clusters, we check whether they have some subsequences that share some parts in common. All clusters sharing such overlaps together can recover the complete picture of the issue. Thus, we regard them as describing an identical issue.
% Engineers can label the type of each issue pattern. 
% and link it to an identified remediation operation. In this way, actionable operations can be quickly recommended for the mitigation of similar failures. %Even though some failures are not typical and tend not to occur again, they can be simply categorized as Others. %and \textit{will not degrade the performance of ADSketch in terms of anomaly detection}, which does not require the existence of similar anomalous patterns.\hy{based on similarity, you can always find the most similar one (even the similarity is low)}
% The above process can be automated by directly linking the found metric patterns and the ongoing issues identified by engineers.
% In real-world scenarios, irrelevant metric patterns may be linked to common issues due to insufficient data or low data quality. Therefore, we leave the final decision to engineers, i.e., they can break the false positives (i.e., mis-clustered patterns) or bridge the false negatives (i.e., non-clustered patterns), which is also a means of incorporating human knowledge. Such human intervention can take effect immediately. % Nevertheless, we believe the confidence provided can serve as a valuable reference.
Finally, for each metric pattern, domain engineers will label the type of performance issue that triggers it. Particularly, one pattern can have multiple labels simultaneously. The metric patterns with overlaps will share the same set of performance issue labels.

%\hy{could change Human Knowledge to "Labeling"? in Figure 2, it seems that "Human Knowledge" is a must component. Also, in experiments need to say how much is the labeling effort? how many are required to be labelled, the small the better. or, can consider if the adaptive learning part is indeed necessary.} \zb{I will revise the framework diagram}

%\hy{also,if we assume that the offline detection results are all correct, then no need this human knowledge input?} \zb{I think I make it a bit confusing, revised now}
% \zb{human labeling for previous encountered patterns; engineers need to identify the failure types and \textcolor{red}{separate mis-clustered patterns}; human knowledge can be immediately incorporated into the model}

\subsection{Online Anomaly Detection}

\subsubsection{Anomaly Detection on the Fly}

Based on the metric patterns identified in Algorithm~\ref{algo:pattern_discovery}, we now describe our algorithm (Algorithm~\ref{algo:anomaly_detection}) for anomaly detection in online scenarios. The idea is straightforward: given a new metric subsequence $t$ with length $m$, we search for its most similar metric pattern (line 1-2) and check which pattern pool it comes from. If $t$ is more similar to an abnormal pattern, it will be predicted as anomalous; otherwise, normal (line 3-7). In real-world systems where monitoring metrics are generated in a stream manner, this process is continuously running for all coming subsequences. When an anomaly is identified, we would like to provide more interpretation about it, e.g., what kinds of performance issues have happened. This is done by simply recommending the issues associated with the most similar metric pattern for all involved metrics. Particularly, in Algorithm~\ref{algo:pattern_discovery}, each cluster (i.e., $C$ at line 6) contains all subsequences that are deemed as similar. The design of our online anomaly detection only requires the mean vector of each cluster, i.e., $\mu_C$. Thus, instead of keeping all its members (which is storage-intensive), the clusters can be simply represented by their mean vectors.

Note that the offline and online anomaly detection can work collaboratively as a performance anomaly detector without the interpretability component, which requires human intervention.

So far the metric patterns for anomaly detection are discovered based on historical data. However, due to the dynamics of online service systems ({e.g., software upgrade, customer behavior change}), the metrics may experience concept drift~\cite{han2020toward,gama2014survey}, which produces brand-new patterns. Thus, an adaptive learning mechanism is desirable to help adapt to such unprecedented patterns and update the metric patterns accordingly. In the next section, we will introduce the algorithm to this end called adaptive pattern learning.

\begin{algorithm}[t]
\caption{Adaptive Pattern Learning}
\label{algo:adaptive_learning}
\normalsize
\SetAlgoLined
\KwIn{$t$, $\mathcal{P}_n$, $\mathcal{P}_a$, $\mu_C$, $\mathcal{S}_C$, and $\mathcal{R}_C$}
\KwOut{Updated variables: $\mathcal{P}_n$, $\mathcal{P}_a$, $\mu_C$, $\mathcal{S}_C$, and $\mathcal{R}_C$}

$\mathcal{D}_t\gets {\rm PairWiseDistance}(t, \mu_C)$

$idx\gets {\rm MinIndex}(\mathcal{D}_t)$

$\mu^{'}\gets (\mu_G[idx]\times \mathcal{S}_C[idx]+t)/(\mathcal{S}_C[idx]+1)$

$d_w\gets {\rm Distance}(\mu_C[idx], \mu^{'})+\mathcal{R}_C[idx]$

$d_t\gets {\rm Distance}(t, \mu^{'})$

$d^{'}\gets {\rm Max}(d_t, d_w)$

$d_n, d_a\gets {\rm Max}(\mathcal{R}_C[\mathcal{P}_n]), {\rm Max}(\mathcal{R}_C[\mathcal{P}_a])$

% $d\gets d_a~{\rm \textbf{if}}~idx\in \mathcal{P}_a~{\rm \textbf{else}}~d_n$

${\rm \textbf{if}}~idx\in \mathcal{P}_a~{\rm \textbf{then}}~d\gets d_a~{\rm \textbf{else}}~d\gets d_n~{\rm \textbf{end}}$

% $d\gets d_n$

% \If{$idx\in \mathcal{P}_a$}{
%     $d\gets d_a$
% }

% $\mathcal{P}_a^{'}\gets {\rm Copy}(\mathcal{P}_a)$

\eIf{$\mathcal{D}_t[idx] < d$}{
\tcp{add $t$ to the most similar cluster}

    % $\mathcal{S}_C[idx]\gets \mathcal{S}_C[idx]+1$, $\mu_C[idx]\gets \mu^{'}$, $\mathcal{R}_C[idx]\gets d^{'}$
    
    $\mu_C[idx], \mathcal{S}_C[idx], \mathcal{R}_C[idx]\gets \mu^{'}, \mathcal{S}_C[idx]+1, d^{'}$
    
    % \not\in \mathcal{P}_a^{'}$
    
    \eIf{$\mathcal{S}_C[idx]>{\rm Max}(\mathcal{S}_C[\mathcal{P}_a])~{\rm and}~idx~is~a~new~cluster$}{
        $\mathcal{P}_n\gets {\rm Append}~\mathcal{P}_n~{\rm with}~idx$
    
        $\mathcal{P}_a\gets {\rm Remove}~idx~{\rm from}~\mathcal{P}_a$
    }{
        % \eIf{$idx\in \mathcal{P}_a$}{
        %     $d_a\gets {\rm Max}(d_a, d^{'})$
        % }{
        %     $d_n\gets {\rm Max}(d_n, d^{'})$
        % }
    
        $d\gets {\rm Max}(d, d^{'})$ \tcp{$d$ will be assigned to $d_n$ or $d_a$ accordingly}
    }
}{
    \tcp{create a new anomalous cluster for $t$}
    $\mathcal{P}_a\gets {\rm Append}~\mathcal{P}_a~{\rm with}~{\rm Length(\mu_C)+1}$
    
    $\mu_G\gets {\rm Append}~\mu_G~{\rm with}~t$
    
    $\mathcal{R}_C\gets {\rm Append}~\mathcal{R}_C~{\rm with}~0$
    
    $\mathcal{S}_C\gets {\rm Append}~\mathcal{S}_C~{\rm with}~1$
}

\end{algorithm}

\begin{figure}[t]
    \centering
    % \vspace{-0.08in}
    \includegraphics[width=0.94\linewidth]{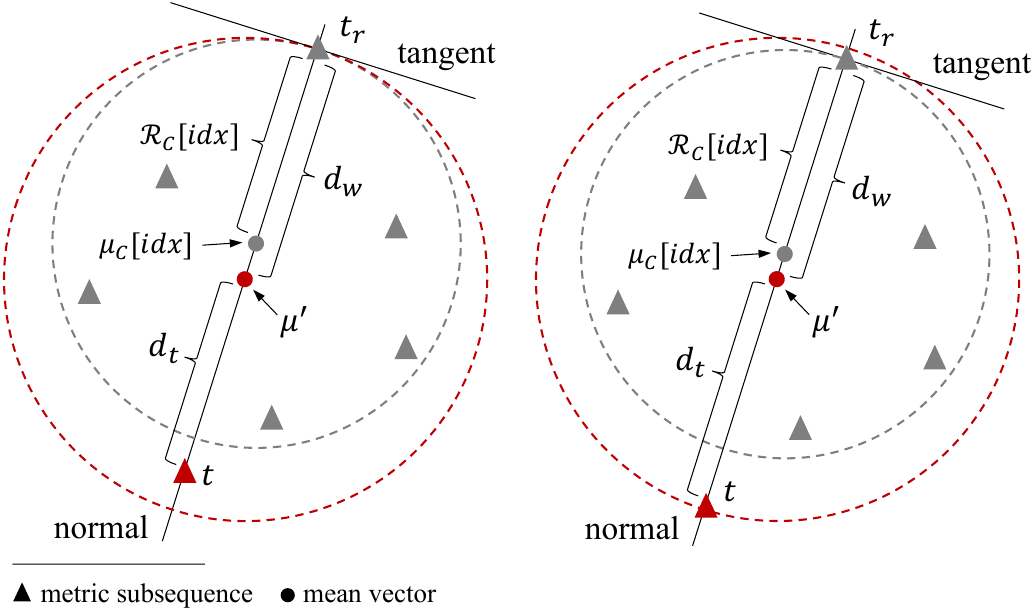}
    % \vspace{-0.1in}
    \caption{The update of the radius of a cluster}
    \label{fig:radius_update}
    % \vspace{-0.12in}
\end{figure}

\subsubsection{Adaptive Pattern Learning}
\label{sec:adaptive_learning}

% \hy{If space is limited, this could be the content of next paper? maybe can focus on offline and online learning in this paper} \zb{I think I can make it regarding the space. The online AD part is actually quite trivial. Considering most existing work are not adaptive, it will make the paper more interesting. Without the adaptive part, I am afraid the contribution is not enough for ICSE.}
The algorithm of adaptive pattern learning is presented in Algorithm~\ref{algo:adaptive_learning}, which automatically updates metric patterns during streaming anomaly detection. To start with, for each cluster, we calculate its size and the maximum distance between its mean vector and all members (which we refer to as \textit{radius}), denoted as $\mathcal{S}_C$ and $\mathcal{R}_C$, respectively. In particular, the size and radius of clusters with only a single member are one and zero. For adaptive pattern learning, all clusters can be sufficiently represented with the following properties: $\mu_C$, $\mathcal{S}_C$, and $\mathcal{R}_C$. All subsequences can be discarded.

The main idea is that given a new subsequence $t$, we determine whether it possesses a known metric pattern carried by an existing cluster. If yes, the cluster will absorb $t$ as a new member and update its properties; otherwise, a brand-new anomalous cluster with only $t$ itself will be created, representing an unseen metric pattern. Specifically, we first search for the closest pattern of $t$ (line 1-2). Then, we determine whether $t$ should become a new member to the corresponding cluster by checking if the distance $\mathcal{D}_t[idx]$ is smaller than the largest radius recorded in all clusters, i.e., $\mathcal{D}_t[idx]\leq {\rm Max}(\mathcal{R}_C)$. If it is the case, $t$ should be considered as an old pattern; otherwise, it should be expressing a new pattern.

When a cluster accepts a new member (line 9-16), we need to update its mean vector $\mu_C[idx]$ (i.e., the metric pattern), size $\mathcal{S}_C[idx]$, and radius $\mathcal{R}_C[idx]$. For $\mu_C[idx]$, it can be precisely updated by the equation at line 3 (i.e., $\mu^{'}$). $\mathcal{S}_C[idx]$ can be trivially updated by increasing itself by one. The update of the radius $\mathcal{R}_C[idx]$ is a bit problematic. We cannot directly calculate the new radius as the original subsequences are not available. To address this problem, we employ the worst-case distance for approximation. As shown in Fig.~\ref{fig:radius_update}, the new radius reaches its maximum value when $t$ lies in the (inward-pointing) normal of the tangent space at the member yielding the radius (denoted as $t_r$)~\cite{boyd2004convex}, which can be calculated by the equation at line 4. We omit the proof, which is standard. Two cases are possible. The first (the left subfigure) is that $t_r$ continues to be the farthest member from the new mean $\mu^{'}$. The second (the right subfigure) is that $t$ takes the place of $t_r$ and becomes the farthest one. Therefore, besides $d_w$, we also compute the distance between $t$ and $\mu^{'}$, i.e., $d_t$, and compare them (line 4-6). The bigger one will be the new radius (line 10). Recall we need to check if $\mathcal{D}_t[idx]\leq {\rm Max}(\mathcal{R}_C)$ to decide whether or not $t$ should be taken as a new member. Considering the high imbalance between normal and abnormal clusters, we maintain two maximum radii for them, denoted as $d_n$ and $d_a$, respectively (line 7).
% We set $d_a=d_n$ if there is no abnormal cluster, i.e., $\mathcal{T}_a$ contains no anomalies.
Once a cluster alters its radius, we reset the maximum radius of its kind ($d_n$ or $d_a$ as determined by line 8) if it is exceeded by $d^{'}$ (line 15). On the other hand, if the cluster rejects $t$, we form a new anomalous cluster containing only $t$ by properly setting its properties (line 18-21). % This is to avoid missing crucial unprecedented metric patterns as the unknown pattern could be either normal or abnormal.

An issue with this strategy is that false positives will accumulate in $\mathcal{P}_a$ as the unseen patterns can also be normal. We alleviate it by setting a threshold to the size of the newly-formed anomalous clusters (line 11). The role of the cluster will be switched from abnormal to normal if its size exceeds the threshold (line 12-13). The rationale is that performance anomalies are generally rare events. A large anomalous cluster would mean the particular type of issue it represents occurs too often. However, a pattern with a large frequency tends to be the metric's normal behavior. In this paper, we simply set the default threshold as the largest size of the anomalous clusters identified in the offline stage, i.e., ${\rm Max}(\mathcal{S}_C[\mathcal{P}_a])$. Nevertheless, more sophisticated strategies can be applied by, for example, considering the distribution of clusters' sizes.

\subsection{Time and Space Complexity}

\subsubsection{Time Complexity}

For Algorithm~\ref{algo:pattern_discovery}, the theoretical time complexity of operation STAMP is $\mathcal{O}(n^2)$. Thus, line 1-2 require $\mathcal{O}(l_n^2)$ and $\mathcal{O}(l_a^2)$, respectively, where $l_n$ and $l_a$ are the length of $\mathcal{T}_n$ and $\mathcal{T}_a$. % where $l_a^2$ is often much larger than $l_n^2$.
Another operation with an interesting time complexity is the affinity propagation algorithm (line 7), whose complexity is quadratic in the number of clusters (which is often small), i.e., $\mathcal{O}(|C|^2)$. Other operations are of trivial linear time complexity, which is also the case for Algorithm~\ref{algo:anomaly_detection} and Algorithm~\ref{algo:adaptive_learning}. Overall, ADSketch owns a time complexity of $\mathcal{O}(n^2)$ ($\mathcal{O}(l_n^2+l_a^2+|C|^2)$). Fortunately, unlike other models such as deep neural networks, STAMP can be embarrassingly parallelized by distributing its unit operation (SPW distance calculation) to multi-core processors~\cite{yeh2016matrix}. Moreover, STAMP has an ultra-fast approximation to generate results in an anytime fashion.

\subsubsection{Space Complexity}

As described in Sec.~\ref{sec:adaptive_learning}, pattern clusters have a lightweight representation, i.e., $\mu_C$, $\mathcal{S}_C$, and $\mathcal{R}_C$. We also need $\mathcal{P}_n$ and $\mathcal{P}_a$ to distinguish anomalous patterns from the normal ones. Besides $\mu_C$ whose space complexity is $\mathcal{O}(m\times |C|)$, other vectors are of $\mathcal{O}(|C|)$. Therefore, the dominant term of space complexity is $\mathcal{O}(m\times |C|)$. Since both $m$ and $|C|$ are usually small, the space overhead of ADSketch can be considered trivial.

\section{Experiments}
\label{sec:exp}

In this section, we evaluate ADSketch using both public data and real-world metric data collected from the industry. Particularly, we aim at answering the following research questions.

\textbf{RQ1}: How effective is ADSketch's offline anomaly detection?

\textbf{RQ2}: How effective is ADSketch's online anomaly detection?

\textbf{RQ3}: How effective is ADSketch's adaptive pattern learning?

% \textbf{RQ4}: How effective is ADSketch's interpretability?
% \zb{Sensitivity to training data with anomalies?}

The evaluation process of much existing work, e.g., ~\cite{su2019robust,ren2019time}, essentially corresponds to the process adopted in RQ1 (i.e., the offline anomaly detection phase), because the threshold they select for anomaly alerting is determined by iterating the full range of its possible values. The best results achieved during the iteration process are reported. To fully examine the performance of different methods in online scenarios, we fix models' data and parameters (including the threshold learned in offline mode) as if they are deployed in production systems, i.e., RQ2. The online adaptability of ADSketch will be evaluated in RQ3.

\subsection{Experiment Setting}

\subsubsection{Dataset}

To evaluate the effectiveness of ADSketch in performance anomaly detection, we conduct experiments on two publicly available datasets. Moreover, to confirm its practical significance, we collect a production dataset from a large-scale online service of Huawei Cloud. Table~\ref{tab:data_statistics} summarizes the statistics of the datasets.

\begin{table}[t]
\begin{center}
\caption{Dataset Statistics}
\label{tab:data_statistics}
\normalsize
 \begin{tabular}{c|c|c|c}
  \toprule
  
  \textbf{Dataset} & \textbf{\#Curves} & \textbf{\#Points} & \textbf{Anomaly Ratio}\\
  
  \midrule
  \midrule

Yahoo & 67 & 94,866 & 1.8\% \\
AIOps18 & 58 & 5,922,913 & 2.26\% \\
Industry & 436 & 4,394,880 & 1.07\% \\

  \bottomrule
 \end{tabular}
\end{center}
\end{table}

\textbf{Public dataset}. The public datasets for experiments are Yahoo~\cite{yahoo} and AIOps18~\cite{aiops18_data,ren2019time}. Particularly, we do not conduct online anomaly detection on Yahoo due to its limited number of anomalies.

\begin{itemize}
    % \item \textit{NYC Taxi}. NYC Taxi dataset contains the New York City (NYC) taxi demand spanning from July, 2014 to July, 2015 with 17,520 timestamps (30 minutes intervals) in total. Each timestamp records the number of observed passengers. This dataset is selected from Numenta Anomaly Benchmark (NAB)~\cite{nab}, a benchmark for the evaluation of anomaly detection algorithms, especially on streaming data. There are five collective anomalies, which are caused by special events, i.e., NYC marathon, Thanksgiving, Christmas, New Years day, and a snow storm~\cite{braei2020anomaly}. Particularly, we select the last 10\% of data as the anomaly-free input and rest as the input containing anomalies.

    % \item \textit{ECGs}. ECGs dataset~\cite{keogh2005hot} contains nine pairs of signal curves. Each signal curve has two parts: one part is anomaly-free and the other has a single anomaly corresponding to a pre-ventricular contraction. We use them as the normal and abnormal input to conduct offline anomaly detection.
    
    \item \textit{Yahoo}. Yahoo released by Yahoo! Research~\cite{yahoo} is a benchmark dataset for time series anomaly detection. Part of the dataset is synthetic (which is simulated by algorithmically injecting anomalies), and part of the dataset is collected from the real traffic of Yahoo services. The anomalies in the real dataset are manually labeled. All time series are sampled every hour. In particular, as our goal is detecting performance anomalies for online services, we only use the real dataset, which reflects the real-world service performance issues. For each time series, we select the first 300 data points as the anomaly-free input (any anomalies are ignored), while the remaining part as the input for offline anomaly detection.

    \item \textit{AIOps18}. AIOps18 dataset was released by an international AIOps competition held in 2018~\cite{aiops18_intro}. The dataset is composed of multiple metric time series collected from the web services of large-scale IT companies. Particularly, the dataset contains two types of metrics, i.e., service metrics and machine metrics. The service metrics record the scale and performance of the web services, including response time, traffic, connection errors; while the machine metrics reflect the health states of physical machines, including CPU usage, network throughput. Some metric time series has a sampling interval of one minute, while that of others is five minutes. Each metric has a training and a testing time series. Thanks to its large quantity, we follow the following procedure to separate the data for ADSketch offline and online anomaly detection. First, we extract a small part of the training time series that is anomaly-free, which often contains thousands of data points. Then, we use the remainder of the training time series for offline anomaly detection. Finally, the whole testing time series will be employed for online anomaly detection. We also compare the performance of online anomaly detection with and without the adaptive learning component.
\end{itemize}

\textbf{Industrial dataset}. To evaluate ADSketch in production scenarios, we collect various metrics (e.g., Application CPU Usage, Interface Throughput, Request Timeout Number, Round-trip Delay) from a large-scale online service (we conceal the name for privacy concern) of Huawei Cloud. The system under study produces millions of metric time series, which contain an abundance of different metric patterns.
% Besides offering traditional services such as Virtual Network, VPN Gateway, it also features intelligent IP networks and other next-generation network solutions.
The number of metric curves collected is 436, which come from multiple instances of virtual machines, containers, and applications of the selected service system. For each metric, we collect one week of data with a sampling interval of one minute, resulting in more than four million data points in total. The anomalies representing the performance issues of the service are labeled by experienced domain engineers. From Table~\ref{tab:data_statistics}, we can see that the anomaly ratio is very low. Particularly, we use the first day as the anomaly-free input, whose anomalies (if any) are simply ignored. The next three days are used for offline anomaly detection. Finally, we conduct online anomaly detection on the remaining three days, where we also evaluate the adaptability of different approaches to unseen anomaly patterns.

\subsubsection{Evaluation Metrics}

As anomaly detection is essentially a binary classification problem, i.e., normal and abnormal, we employ \textit{precision}, \textit{recall}, and \textit{F1 score} for evaluation. They can gauge the performance of an anomaly detection algorithm at a fine-grained level. A satisfactory algorithm should be able to quickly and precisely detect both the occurrence and duration of performance anomalies. Specifically, precision measures the percentage of anomalous metric points that are successfully identified as anomalies over all the metric points that are predicted as anomalous: $precision=\frac{TP}{TP+FP}$. Recall calculates the portion of anomalous metric points that are successfully identified by ADSketch over all the actual anomalous points: $recall=\frac{TP}{TP+FN}$. Finally, the F1 score is the harmonic mean of precision and recall: $F1~Score=\frac{2\times precision\times recall}{precision+recall}$. $TP$ is the number of anomalous metric points that are correctly discovered by ADSketch; $FP$ is the number of normal metric points that are wrongly predicted as an anomaly by ADSketch; $FN$ is the number of anomalous metric points that ADSketch fails to notice. Since there are multiple metrics in each dataset, we report their average weighted by the size of each metric time series.

% \hy{need to describe the Average F1 and W. F1 used in Tables 3 and 4,5}

% For example, 
% minutes of service downtime could become an expensive drain on company revenue and user dissatisfaction~\cite{chen2020towards}, while seconds of delay in Electrocardiograms could cause irreparable consequences to patients' lives. %How long the anomalous states should last is also an important factor affecting doctors' judgement.
% \hy{how do you handle the "how long" issue then?} \zb{The strategy is introduced in the second step: Failure Pattern Identification. But I have no experiments to quantitatively show it, only a case study}\hy{perhaps not to highlight it...}

\subsubsection{Comparative Methods}

The following methods are selected for comparative evaluation of ADSketch. As all baselines have open-sourced their code, we directly borrow the implementations and follow the procedure of model training and parameter tuning introduced in each method.
% Our code and sample data are publicly available on GitHub\footnote{We will open source them after double-blind review}.

\begin{itemize}
    \item \textit{LSTM}~\cite{hundman2018detecting,zhao2021predicting}. This method employs Long Short-Term Memory (LSTM) network to capture the normal behaviors of metrics in a forecasting-based manner. Specifically, it predicts the next values of a metric based on its past observations. The predicted values are then compared with the actual values. Anomaly warnings will be raised if the differences exceed the pre-defined thresholds.
    
    \item \textit{Donut}~\cite{xu2018unsupervised}. Donut adopts the Variational Autoencoder (VAE) framework to properly reconstruct the normal metric subsequences. The trained model will have a large reconstruction loss when it meets anomalous instances, which serves as the signal to alert anomalies.
    
    \item \textit{LSTM-VAE}~\cite{park2018multimodal}. Similar to Donut, this work detects anomalies based on metric subsequence reconstruction. It combines LSTM and VAE in the model design.
    
    \item \textit{LODA}~\cite{pevny2016loda}. LODA is an online anomaly detector based on the ensemble of a series of one-dimensional histograms. Each histogram approximates the probability density of input data projected onto a single projection vector. LODA calculates the likelihood of an anomaly based on the joint probability of the projections.
    
    \item \textit{iForest}~\cite{liu2008isolation}. Isolation Forest (iForest) is composed of a collection of isolation trees, which isolates anomalies based on random subsets of the input features. The height of an input sample, averaged over the trees, is a measure of its normality. Samples with noticeably shorter heights are likely to be anomalies. We use metric subsequences as the input samples.
    
    \item \textit{DAGMM}~\cite{zong2018deep}. DAGMM utilizes a deep autoencoder to generate a low-dimensional representation for each input data point, which is further fed into a Gaussian Mixture Model to estimate the anomaly score.
    
    \item \textit{SR-CNN}~\cite{ren2019time}. SR-CNN first applies Spectral Residual to highlight the most important regions for seasonal metric data where anomalies often reside. It then trains a Convolutional Neural Network (CNN) through synthetic anomalies to detect the real anomalies.
\end{itemize}

\subsection{Experimental Results}

% In this section, we conduct experiments to answer the research questions.

% \textcolor{red}{Particularly, $m$ should be roughly equal to the size of small sets of consecutive anomalies (not necessarily the smallest one). $p$ should roughly equal to the ratio of anomalies in a time series. The best setting is often reached when precision and recall have comparable values.}

\begin{table*}[t]
\begin{center}
\caption{Experimental Results of Offline Anomaly Detection}
% \hy{what are avgF1 and W.F1, need describe} \zb{mentioned at the begining of RQ1}
\label{tab:offline_res}
\normalsize
 \begin{tabular}{c|c|c|c|c|c|c|c|c|c}
  \toprule
  & \multicolumn{3}{c|}{Yahoo} & \multicolumn{3}{c|}{AIOps18} & \multicolumn{3}{c}{Industry} \\
  
  \textbf{Method} & \textbf{precision} & \textbf{recall} & \textbf{F1 score} & \textbf{precision} & \textbf{recall} & \textbf{F1 score} & \textbf{precision} & \textbf{recall} & \textbf{F1 score}\\
  
  \midrule
  \midrule
  
LSTM & 0.598 & \textbf{0.706} & 0.530 & 0.499 & 0.531 & 0.518 & 0.704 & 0.656 & 0.632 \\
LSTM-VAE & 0.622 & 0.634 & 0.484 & 0.510 & 0.625 & 0.537 & 0.717 & 0.639 & 0.622 \\
Donut & 0.530 & 0.658 & 0.524 & 0.405 & 0.527 & 0.382 & 0.693 & 0.628 & 0.604 \\
LODA & \textbf{0.754} & 0.583 & 0.428 & 0.553 & 0.429 & 0.401 & 0.583 & 0.498 & 0.529 \\
iForest & 0.713 & 0.597 & 0.437 & 0.555 & 0.439 & 0.413 & 0.616 & 0.567 & 0.538 \\
DAGMM & 0.643 & 0.517 & 0.401 & 0.590 & 0.477 & 0.461 & 0.597 & 0.542 & 0.530 \\
SR-CNN & 0.433 & 0.618 & 0.307 & 0.424 & 0.387 & 0.363 & 0.519 & 0.471 & 0.434 \\
ADSketch & 0.511 & 0.673 & \textbf{0.541} & \textbf{0.744} & \textbf{0.670} & \textbf{0.677} & \textbf{0.811} & \textbf{0.813} & \textbf{0.740} \\

  \bottomrule
 \end{tabular} 
\end{center}
\end{table*}

% \begin{table*}[t]
% \begin{center}
% \caption{Experimental Results of Offline Anomaly Detection on NYC Taxi and ECGs Datasets}
% \label{tab:offline_res}
% \normalsize
%  \begin{tabular}{c|c|c|c|c|c|c}
%   \toprule
%   & \multicolumn{3}{c|}{NYC Taxi} & \multicolumn{3}{c}{ECGs}\\
  
%   \textbf{Methods} & \textbf{Precision} & \textbf{Recall} & \textbf{F1 Score} & \textbf{Precision} & \textbf{Recall} & \textbf{F1 Score} \\
  
%   \midrule
%   \midrule

% LSTM-NDT & 0.985 & 0.261 & 0.413 & 0.521 & 0.204 & 0.246 \\
% LSTM-VAE & \textbf{0.997} & 0.281 & 0.438 & 0.672 & 0.207 & 0.249 \\
% Donut & 0.311 & 0.466 & 0.373 & 0.615 & \textbf{0.448} & 0.357 \\
% LODA & 0.639 & 0.182 & 0.283 & 0.634 & 0.178 & 0.241 \\
% iforest & 0.295 & 0.101 & 0.150 & 0.595 & 0.211 & 0.262 \\
% DAGMM & 0.212 & 0.199 & 0.205 & \textbf{0.693} & 0.204 & 0.238 \\
% SR-CNN & 0.186 & 0.127 & 0.151 & 0.649 & 0.260 & 0.219 \\
% ADSketch & 0.576 & \textbf{0.808} & \textbf{0.672} & 0.520 & 0.440 & \textbf{0.456} \\

%   \bottomrule
%  \end{tabular} 
% \end{center}
% \end{table*}

\subsubsection{\textbf{RQ1} The Effectiveness of ADSketch's Offline Anomaly Detection}

To answer this research question, we compare ADSketch with the baselines in the offline setting.
% Except for NYC Taxi which contains only one time series, we average the results of different time series for other datasets. The models are first trained by consuming the anomaly-free time series. Then, they are evaluated on the one that contains anomalies. In this step, we keep the tuning strategy adopted by the respective papers such as threshold iteration. Particularly, in ADSketch, all data points that appear in any of the anomalous clusters are predicted as anomaly.
% \hy{if space allows, briefly introduce each compared methods (e..g, using bulletin points}
The results are shown in Table~\ref{tab:offline_res}, where we can see the average F1 score of ADSketch outperforms all baseline methods in all datasets. In AIOps18 and Industry, the improvement achieved by ADSketch is more significant. In particular, the patterns of anomalies in Yahoo are relatively simple. By iterating over all possible values of the anomaly threshold, the baselines can find the best setting for the dataset under study. Among them, LSTM~\cite{hundman2018detecting,zhao2021predicting} and Donut~\cite{xu2018unsupervised} achieve comparable performance compared to that of ADSketch (i.e., 0.541), whose average F1 scores are 0.53 and 0.524, respectively. Moreover, LSTM~\cite{hundman2018detecting,zhao2021predicting} has the best recall (i.e., 0.706), while the best precision (i.e., 0.754) goes to LODA~\cite{pevny2016loda}. DAGMM and SR-CNN turn out to be the worst methods in this dataset.
% In NYC Taxi dataset, ADSketch performs exceptionally well, while others struggle the most, especially the recall. By inspecting the data, we found this is because the anomalies are composed of a collection of anomalous points, i.e., a special event period. Other models (i.e., LSTM-NDT, LSTM-VAE, and LODA) lack the ability to explicitly detect anomalies in the level of KPI subsequences. Thus, they only manage to locate the most deviated data points, leading to a high precision (e.g., LSTM-VAE has the best precision) and yet an extremely low recall. The remaining models perform poorly in this dataset. Similarly results can be observed in ECGs. This dataset also contains anomalies that last for some period of time, i.e., an abnormal heart beat. Nevertheless, as ECGs has more obvious anomalous patterns than NYC Taxi does, other models achieve a comparable performance (the best precision and recall are achieved by DAGMM and Donut, respectively).
In terms of AIOps18 and Industry datasets, we can see ADSketch surpasses the baselines by a larger margin. Specifically, the average F1 score of ADSketch in AIOps18 is 0.677, while that of the second-best method (i.e., LSTM-VAE) is 0.537. ADSketch also attains the best precision and recall. In AIOps18, the anomaly patterns are much more complicated. Baselines tend to predict more data points as anomalous, leading to a lower precision. Different from them, ADSketch is able to precisely capture them and outperforms other methods. The situation is similar in Industry. % One important reason is that AIOps18 provides sufficient data for models to learn from.
Particularly, this dataset is collected from online services, and many of its metric curves possess more perceivable and regular patterns. Thus, all methods perform better in this dataset than in the other two. The average F1 scores of ADSketch and the second-best method (i.e., LSTM) are 0.740 and 0.632, respectively.

In Table~\ref{tab:offline_res}, we can see among all comparative methods, LSTM and LSTM-VAE have better overall performance, which are forecasting-based and reconstruction-based methods, respectively. They both try to model the normal patterns of a metric time series and alert anomalies once the metric significantly deviates from the learned patterns. The difference is that a forecasting-based method aims to predict the next metric values and a reconstruction-based method tries to encode and regenerate metric subsequences. We can see except for LSTM-VAE in Yahoo, these two methods attain the best results compared to other baseline counterparts in the other two datasets. However, LSTM lacks the ability to explicitly detect anomalies in the level of subsequence. Many anomalies are composed of a collection of anomalous points corresponding to the period of performance issues. LSTM-VAE does not take into account the relationship among subsequences. Many suspicious subsequences are not necessarily anomalies if they often occur in the history of the service systems. Compared to them, ADSketch is able to simultaneously learn the subsequence-level features and consider the context of metric time series.

\begin{table}[t]
\begin{center}
\caption{Experimental Results of Online Anomaly Detection}
\label{tab:online_res}
\normalsize
 \begin{tabular}{c|c|c|c|c|c|c}
  \toprule
  & \multicolumn{3}{c|}{AIOps18} & \multicolumn{3}{c}{Industry} \\
  
  \textbf{Method} & \textbf{prec.} & \textbf{rec.} & \textbf{F1} & \textbf{prec.} & \textbf{rec.} & \textbf{F1}\\
  
  \midrule
  \midrule
  
LSTM & 0.425 & 0.462 & 0.408 & 0.612 & 0.606 & 0.592 \\
LSTM-VAE & 0.336 & 0.521 & 0.389 & 0.624 & 0.598 & 0.601 \\
Donut & 0.431 & 0.326 & 0.376 & 0.662 & 0.581 & 0.590 \\
LODA & 0.407 & 0.397 & 0.355 & 0.653 & 0.526 & 0.503 \\
iForest & 0.397 & 0.334 & 0.322 & 0.576 & 0.507 & 0.487 \\
DAGMM & 0.392 & 0.367 & 0.378 & 0.557 & 0.538 & 0.502 \\
SR-CNN & 0.329 & 0.288 & 0.307 & 0.438 & 0.422 & 0.410 \\
ADSketch & \textbf{0.543} & \textbf{0.575} & \textbf{0.507} & \textbf{0.705} & \textbf{0.603} & \textbf{0.606} \\

  \bottomrule
 \end{tabular} 
\end{center}
\end{table}

% \hy{the online detection results for AIOPs dataset are not very good. Perhaps can consider:
% 1) use other datasets; 2) use more metrics such as ROC, Balance, Sensitivity, Fbeta-Measure, TPR, FPR,  etc... This is a highly imbalanced dataset, so not sure of P/R/F metrics for the abnormal class only are sufficient. }

\begin{table}[t]
\begin{center}
\caption{Experimental Results of Adaptive Pattern Learning}
\label{tab:adaptive_res}
\normalsize
 \begin{tabular}{c|c|c|c|c|c|c}
  \toprule
  & \multicolumn{3}{c|}{AIOps18} & \multicolumn{3}{c}{Industry} \\
  
  \textbf{Method} & \textbf{prec.} & \textbf{rec.} & \textbf{F1} & \textbf{prec.} & \textbf{rec.} & \textbf{F1}\\
  
  \midrule
  \midrule

LODA & 0.424 & 0.405 & 0.387 & 0.623 & 0.512 & 0.548 \\
ADSketch & \textbf{0.594} & \textbf{0.557} & \textbf{0.548} & \textbf{0.882} & \textbf{0.856} & \textbf{0.832} \\

  \bottomrule
 \end{tabular}
\end{center}
\end{table}

\subsubsection{\textbf{RQ2} The Effectiveness of ADSketch's Online Anomaly Detection}
% \hy{say why Yahoo is not included} \zb{added to the intro of Public dataset}
We also compare ADSketch against the selected methods for online anomaly detection. Table~\ref{tab:online_res} presents the experimental results. Except for Donut in AIOps18, all models and algorithms encounter an obvious performance degradation in both datasets. Nevertheless, ADSketch manages to maintain the best ranking (0.507 in AIOps18 and 0.606 in Industry), which is followed by LSTM (0.408 in AIOps18) and LSTM-VAE (0.601 in Industry). Particularly, in AIOps18, the average F1 score of different methods drops by 11\%-27\%. This observation demonstrates the existence of unprecedented metric patterns in online scenarios. By relying on the "outdated" data and parameters (e.g., ADSketch's metric patterns and baselines' anomaly thresholds) learned from the offline stage, the methods cannot accommodate them. In addition, by plotting the metric time series, we observe the emergence of concept drift on metrics. This can be caused by software upgrades or the integration of new service components (e.g., virtual machines, containers). In the industrial dataset, the evaluation results of the baselines are more promising (i.e., the average F1 score drops by less than 10\%). This is because the anomalies are triggered by real-world performance issues. The issues have a more natural distribution, and the collected metrics exhibit relatively stable patterns. ADSketch presents a significant performance degradation. We found it is because in some cases, the two metric time series fed to the offline stage are often both anomaly-free. Consequently, no abnormal patterns will be learned, disabling ADSketch to detect anomalies in the online stage. Therefore, when designing an anomaly detection algorithm, adaptability is indispensable.

\subsubsection{\textbf{RQ3} The Effectiveness of ADSketch's Adaptive Pattern Learning}
%\hy{what do the results mean? also, say the differences between RQ2 and RQ3, or can just use one of them? ALso the performance on AIOps is bad, even it requires human input.} \zb{RQ2 detects anomalies without adapting to new patterns. RQ3 detects anomalies and adapts to unseen patterns, hence achieves a better result. RQ3 does not require human input. I have revised section 3.2.2 to make it clearer.}\hy{ok, but the performance of adaptive is not very good. Figure 2 shows that adaptive method requires human labeling...is Figure 2 wrong?} \zb{I use weighted F1 score, which gives better results for AIOps18.}
This research question looks into the issue of online adaptability. Particularly, we only compare ADSketch with LODA, which is the only baseline method with the design of online learning. Similar to RQ2, we only conduct experiments with AIOps18 and Industry datasets. Table~\ref{tab:adaptive_res} shows the experimental results, where we can see ADSketch's adaptive pattern learning indeed brings performance gains. With more anomalous patterns identified, ADSketch is able to detect anomalies more accurately, i.e., a better precision (0.594 in AIOps18 and 0.882 in Industry). The average F1 score also enjoys some improvements, i.e., 0.548 in AIOps18 and 0.832 in Industry. Particularly, in the industrial case, adaptive ADSketch achieves a performance of over 0.8 in all evaluation metrics (even in some cases without any abnormal patterns learned from the offline stage). Such an achievement indicates its potential to meet the industrial requirements of performance anomaly detection. On the other hand, the online version of LODA does not show much performance improvement (i.e., an average F1 score of 0.387 in AIOps18 and 0.548 in Industry), which even falls behind some methods without the capability of online learning.

\subsubsection{Parameter Sensitivity}

\begin{figure}[t]
    \centering
    \includegraphics[width=1.0\linewidth]{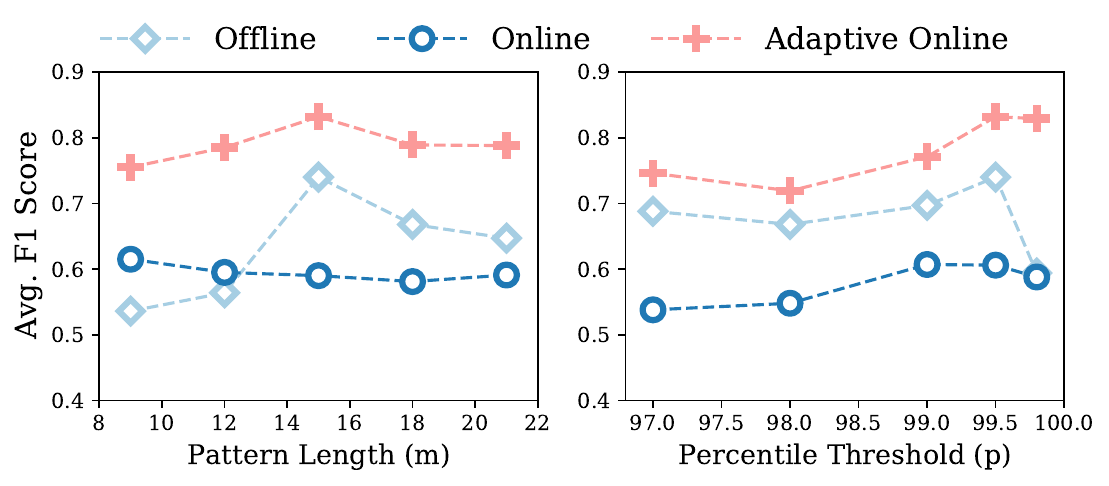}
    \caption{Parameter Sensitivity}
    \label{fig:parameter_sensitivity}
    % \vspace{-0.2in}
\end{figure}

In ADSketch, there are only two parameters to tune (both in Algorithm~\ref{algo:pattern_discovery}), i.e., the pattern length $m$ and the percentile threshold $p$ for identifying deviated metric subsequences. We evaluate the sensitivity of ADSketch to these two parameters by conducting experiments with different settings. Due to space limitations, we only show the results of the Industry dataset. The default value of $m$ and $p$ for the dataset is 15 and 99.5th, respectively. We fix one parameter and employ a different setting for the other one. Specifically, $m$ ranges from 9 to 21, and $p$ varies from 97th to 99.8th. Fig.~\ref{fig:parameter_sensitivity} presents the results. Performance degradation is observed in both offline and online stages when the two parameters deviate from their default setting. The offline stage exhibits a greater sensitivity, and thus, less anomalous metric patterns are captured. Nevertheless, both the online anomaly detection and adaptive pattern learning algorithms achieve stable performance with a smaller set of abnormal patterns. This further confirms ADSketch's capability of new pattern discovery.

%\hy{where? maybe this result is not necessary, if you have more important contents...Can put the results in webpage/supplementary...} \zb{One ASE reviewer mentioned this issue. I will move it to somewhere else if the space doesn't permit.}
%\hy{I think he just tried to find a reason to reject it. The root reason is not this}

\section{Industrial Practice}
\label{sec:discussion}

\subsection{Online Deployment}
%\hy{say more about Success Stories.}
Since October 2020, ADSketch has been successfully incorporated into the performance anomaly detection system of a large-scale online service system in Huawei Cloud. The deployment process can be easily done by leveraging the existing data analytics pipeline, for example, data consumption by Apache Kafka~\cite{kafka}, and online parallel execution by Apache Flink~\cite{flink}. After months of usage, ADSketch has demonstrated its effectiveness on metric-based system troubleshooting. A lot of positive feedback has been received from on-site engineers. Particularly, engineers confirmed its superiority in anomaly detection over the current algorithms (e.g., fixed thresholding, moving average) in operation. One typical case is multiple benign spikes arriving suddenly and consecutively. ADSketch is able to quickly figure out that such recurrent spikes have happened before, which reduces the number of false alerts. In terms of issue understanding, engineers benefited from ADSketch by having readily-available descriptions about the anomaly symptoms. Therefore, we have initialized a project of metric pattern database construction. ADSketch is continuously accumulating anomalous patterns in the database. Moreover, engineers also expressed the need for metric pattern auto-correlation across different metrics. This is because multiple anomalies collectively could constitute a stronger performance issue indicator. We leave the identification of such correlations to our future work.
%\hy{any quantitative numbers? for example, how many problems by solved by ADSketch, the benefits, etc}

\subsection{Case Study}

\begin{figure}[t]
    \centering
    \includegraphics[width=0.9\linewidth]{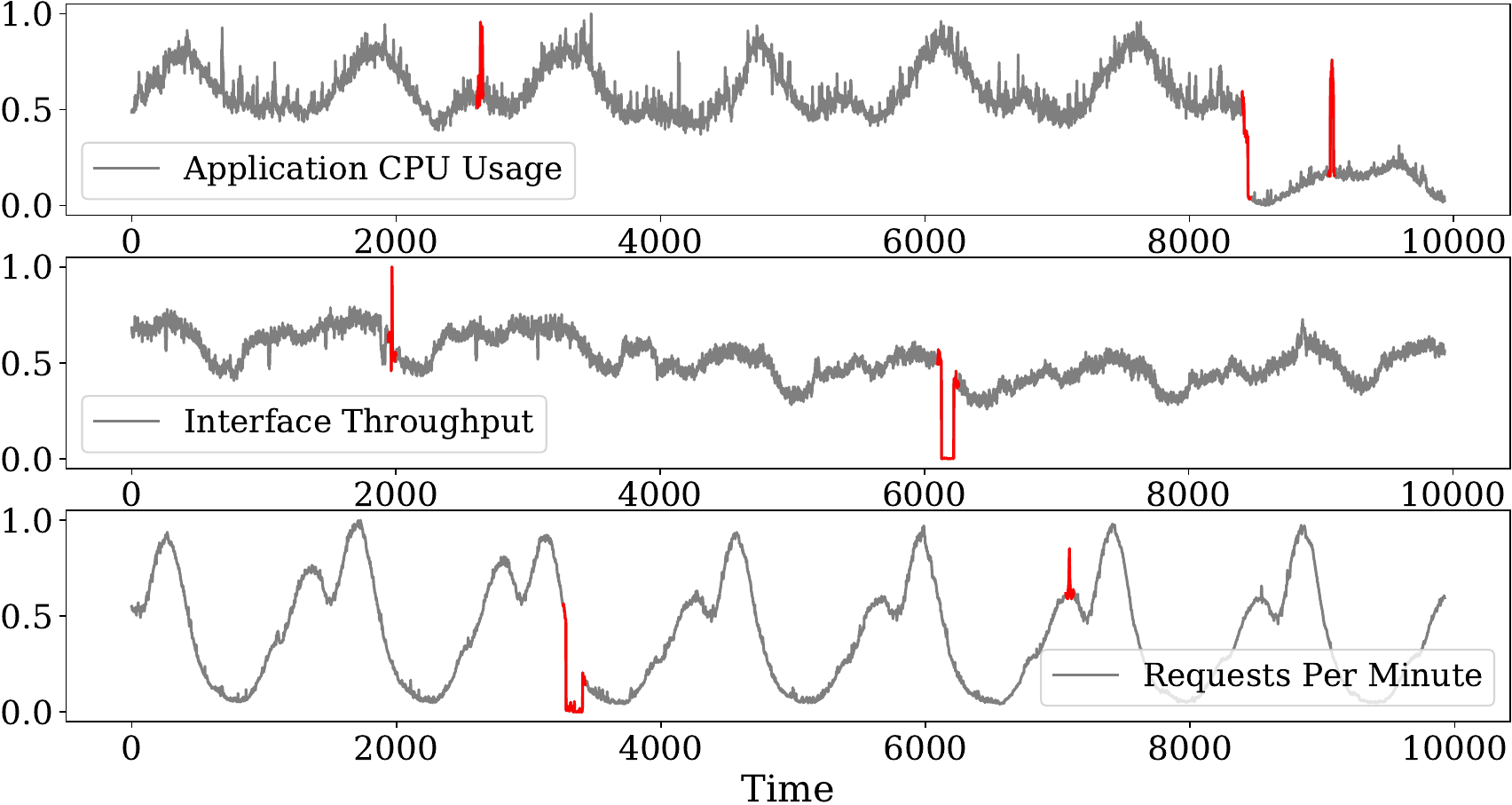}
    % \vspace{-0.05in}
    \caption{Case Study of ADSketch}
    \label{fig:case_study}
    % \vspace{-0.15in}
\end{figure}

We provide some case studies of ADSketch collected from production systems in Fig.~\ref{fig:case_study}, where anomalies are indicated by the red lines. Due to space limitations, we only showcase three metric time series.
% For privacy protection, we erase the name as well as the time of the metric, and apply normalization.
Clearly, all anomalous metric patterns have been successfully located regardless of shape, scale, and length. Each metric time series possesses at least two types of anomalous patterns, e.g., level shifts and spikes. Interestingly, we found the depression in the second metric can help catch a similar pattern in the third metric, demonstrating the feasibility of cross-metric pattern sharing. Moreover, engineers confirmed that these patterns are typical, based on which they can make a good guess about the ongoing issues. For example, the spikes often come from user request surge or network attack; the depressions in the second and third metrics often indicate service restart or link flap. To quantify the interpretability of ADSketch, we label the recurrent performance issues and employ the learned metric patterns to identify them. As performance issues may contain uncertainty~\cite{trubiani2018performance}, we allow one pattern to be associated with multiple labels simultaneously (Sec.~\ref{sec:metric_pattern_interpretability}). During the evaluation, an anomaly interpretation is considered correct if the predicted performance issue appears in the label set. In our experiments, ADSketch attains a promising F1 score of 0.825. This demonstrates the potentials of ADSketch in providing interpretable results to engineers, which can greatly accelerate the investigation of service performance issues.

\subsection{Threats to Validity}
\label{sec:threats_to_validity}

We have identified the following major threats to validity.

\textbf{Internal threats.} The implementation and parameter selection are two critical internal threats to the validity. To reduce the implementation threat, we directly borrow the codes released by the baseline approaches. For the proposed approach, we employ peer code review, i.e., the authors are invited to carefully check the implementation for mistakes. In terms of parameter selection, we conduct multiple comparative experiments with different parameters for all methods. We choose the parameter settings empirically based on the best results.

% (refers to the degree of confidence that the causal relationship being tested is trustworthy and not influenced by other factors or variables.)

\textbf{External threats.} The selection of the service system and the baselines are two main external threats to validity. We choose a large-scale online service of Huawei Cloud, which produces millions of metrics with diverse patterns. Moreover, we detect anomalies by following the basic definition of an anomaly, i.e., the data point that deviates from the majority in a dataset. Thus, ADSketch is generalizable to other systems. For baselines, we select the representative ones in the literature, covering a wide spectrum of techniques.

% (refers to the extent to which results from a study can be applied (generalized) to other situations, groups or events.)

\textbf{Construct threats.} The main construct threat to validity is that the anomaly-free input (i.e., $\mathcal{T}_n$) to Algorithm~\ref{algo:pattern_discovery} actually contains anomalies. Although anomaly-free data are easily obtainable in reality, false negatives could happen if the data are contaminated. We alleviate this issue by applying percentile thresholding to $\mathcal{T}_n$. Specifically, after obtaining the closest subsequence pairs in $\mathcal{T}_n$, we break the connection between those having a distance above the percentile threshold. Thus, the set of anomaly candidates, i.e., $N_i$, becomes larger. If $\mathcal{T}_n$ is indeed clean, this operation is harmless as the (isolated) normal metric subsequences can be grouped with other similar ones again; if not, they will stay isolated and eventually be recognized as anomalies. We have also conducted experiments on some cases where $\mathcal{T}_n$ contains anomalies, and the results show its effectiveness.
\section{Related Work}
\label{sec:related_work}
% \hy{this section is too Data Mining focused, can say more SE-related work, as well as AIOps...can also shorten it a bit}
% a univariate time series anomaly detection survey~\cite{braei2020anomaly}
Performance anomaly detection on time series has been a hot topic. 
Monitoring metrics used to profile the runtime status of a system are usually denoted as multiple univariate time series.
In the literature, anomaly detection methods on time series can be categorized into statistical, traditional machine learning, and deep learning approaches.
In industry, Autoregressive Moving Average Model (ARMA)~\cite{chambers2012discrete} remains the most popular statistical method to detect obvious anomalous data points from univariate time series.
To capture complex anomalous patterns, Ma et al.~\cite{ma2020diagnosing} summarized several type-oriented patterns from the metrics of cloud databases to diagnose the performance degradation in associated online services.

More complex pattern recognition methods utilize machine learning based models.
For example, unsupervised clustering methods can be used to detect anomalous points in time-series data. Similar to our work, Pang et al.~\cite{pang2015lesinn} proposed a clustering-based statistical model called LeSiNN to detect anomaly patterns from history. However, it is not robust in real industry practices due to complicated parameter tuning. With the assumption that anomalous data should be in smaller numbers and isolated from a large number of normal observations, Isolation Forest (iForest)~\cite{liu2008isolation} employs multiple binary trees to distinguish anomalies in non-linear space. Extreme Value Theory (EVT)~\cite{siffer2017anomaly} learns the hidden state of a random variable around the tails of its distribution to adaptively enhance the performance of many statistical and machine learning methods. However, EVT heavily relies on hyperparameter tuning.
%LODA~\cite{pevny2016loda},

In recent years, there has been an explosion of interest in applying neural networks to conduct anomaly detection on time-series data.
For example, Zong et al.~\cite{zong2018deep} proposed a deep autoencoding Gaussian mixture model (DAGMM) to detect anomalous data points from each observed data without considering the temporal dependencies in time series.
To detect complex anomalies in spacecraft monitoring systems, LSTM-NDT~\cite{hundman2018detecting} leverages Long Short-Term Memory (LSTM) networks with nonparametric dynamic thresholding to pursue interpretability throughout the systems.
Zhao et al.~\cite{zhao2021predicting} and Lin et al.~\cite{lin2018predicting} also employed LSTM to predict performance anomalies in software systems.
Inspired by the Spectral Residual algorithm in other domains, Ren et al.~\cite{ren2019time} proposed SR-CNN to detect anomalies from seasonal metric data for large-scale cloud services, which contain the periodic recurrence of fluctuations. 
DONUT~\cite{xu2018unsupervised} designs an unsupervised anomaly detection method based on the Variational Auto-Encoder (VAE) framework to detect anomalies from low-qualified seasonal metric time series with various patterns.
DONUT provides a theoretical explanation compared to other deep learning methods.
LSTM-VAE~\cite{park2018multimodal} combines LSTM networks and the VAE framework to reconstruct the probability distribution of observed data in time series. However, LSTM-VAE ignores the temporal dependencies in time series.
OmniAnomaly~\cite{su2019robust} learns the normal patterns using a large collection of historical data. The anomalous patterns are located from the large margin of reconstruction loss to the normal patterns. 
However, the aforementioned deep learning-based methods usually follow an end-to-end style and play as a black box inside. Due to poor interpretability, the detection results cannot provide engineers with actionable suggestions for fault diagnosis.
% \hy{still not very convicing as hardware should not be a major issue for a Cloud provider, also it can be trained offline (not online)...maybe can say that the DL model does not have the interpretability, so they cannot provide actionable suggestions for fault diagnosis...}
% \yx{fixed}
Furthermore, all these methods %are lack of adaptation 
have difficulties handling unseen metric patterns brought by the frequent updates of online services.

% unsupervised and adaptive anomaly detection
% all previous method no interpretability
% the lack of an online learning

% handcrafted features in previous work

% failure profiling or failure sketching

% failure identification and diagnosis based  on  KPI in online service systems

% Failure Sketching: A Technique for Automated Root Cause Diagnosis of In-Production Failures
\section{Conclusion}
\label{sec:conclusion}

In this paper, we propose ADSketch, a performance anomaly detection approach based on pattern sketching. By extracting normal and abnormal patterns from metric time series, anomalies can be quickly detected through a comparison with the identified patterns. By associating metric patterns with typical performance issues, ADSketch can provide interpretable results when any known patterns appear again. Moreover, we design an adaptive learning algorithm to help ADSketch embrace unprecedented metric patterns during online anomaly detection. We have conducted experiments on two public datasets and one production dataset collected from a representative online service system of Huawei Cloud. For offline anomaly detection where models' parameters are still being tuned, ADSketch has achieved the highest F1 score, outperforming the existing methods by a significant margin. For online anomaly detection where models are fixed, ADSketch safeguards its best rankings. Finally, the adaptive pattern learning brings noticeable performance gains, especially in the industrial dataset.
% \hy{not on AIOps18 data}
% Feedback from engineers \hy{say "our industrial practice" instead of feedback} confirms its practical benefits conveyed to company X.
From our industrial practice, we have witnessed it shedding light on accurate and interpretable performance anomaly detection, which confirms its practical benefits conveyed to Huawei Cloud. We believe ADSketch is able to assist engineers in service failure understanding and diagnosis.

For future work, we will extend our algorithms to multivariate metric time series. We will also try to provide more detailed information about failures by exploring the correlations among the metric patterns.

% Our source code and sample experimental data are publicly available at : ... \hy{better to release some sample code and data} \zb{I mention it in the contributions.}


\begin{acks}

The work was supported by the Guangdong Key Research Program (No. 2020B010165002), the Research Grants Council of the Hong Kong Special Administrative Region, China (CUHK 14210920), and Australian Research Council (ARC) Discovery Projects (DP200102940 and DP220103044).

% We would like to thank anonymous reviewers for their insightful and constructive comments, which significantly improve the quality of this paper.

\end{acks}

\newpage

\bibliographystyle{ACM-Reference-Format}
\balance
\bibliography{bibliography}


\begin{thebibliography}{39}

%%% the Web address.

\ifx \showCODEN    \undefined \def \showCODEN     #1{\unskip}     \fi
\ifx \showDOI      \undefined \def \showDOI       #1{#1}\fi
\ifx \showISBNx    \undefined \def \showISBNx     #1{\unskip}     \fi
\ifx \showISBNxiii \undefined \def \showISBNxiii  #1{\unskip}     \fi
\ifx \showISSN     \undefined \def \showISSN      #1{\unskip}     \fi
\ifx \showLCCN     \undefined \def \showLCCN      #1{\unskip}     \fi
\ifx \shownote     \undefined \def \shownote      #1{#1}          \fi
\ifx \showarticletitle \undefined \def \showarticletitle #1{#1}   \fi
\ifx \showURL      \undefined \def \showURL       {\relax}        \fi
% invisible to TeX
\providecommand\bibfield[2]{#2}
\providecommand\bibinfo[2]{#2}
\providecommand\natexlab[1]{#1}
\providecommand\showeprint[2][]{arXiv:#2}

\bibitem[\protect\citeauthoryear{??}{aio}{2018a}]%
        {aiops18_intro}
 \bibinfo{year}{2018}\natexlab{a}.
\newblock \bibinfo{title}{KPI Anomaly Detection Competition}.
\newblock
\newblock
\urldef\tempurl%
\url{http://iops.ai/competition_detail/?competition_id=5&flag=1}
\showURL{%
Retrieved April, 2021 from \tempurl}


\bibitem[\protect\citeauthoryear{??}{aio}{2018b}]%
        {aiops18_data}
 \bibinfo{year}{2018}\natexlab{b}.
\newblock \bibinfo{title}{KPI Anomaly Detection Dataset}.
\newblock
\newblock
\urldef\tempurl%
\url{http://iops.ai/dataset_detail/?id=10}
\showURL{%
Retrieved April, 2021 from \tempurl}


\bibitem[\protect\citeauthoryear{Boyd, Boyd, and Vandenberghe}{Boyd
  et~al\mbox{.}}{2004}]%
        {boyd2004convex}
\bibfield{author}{\bibinfo{person}{Stephen Boyd}, \bibinfo{person}{Stephen~P
  Boyd}, {and} \bibinfo{person}{Lieven Vandenberghe}.}
  \bibinfo{year}{2004}\natexlab{}.
\newblock \bibinfo{booktitle}{\emph{Convex optimization}}.
\newblock \bibinfo{publisher}{Cambridge university press}.
\newblock


\bibitem[\protect\citeauthoryear{Brand{\'o}n, Sol{\'e}, Hu{\'e}lamo, Solans,
  P{\'e}rez, and Munt{\'e}s-Mulero}{Brand{\'o}n et~al\mbox{.}}{2020}]%
        {brandon2020graph}
\bibfield{author}{\bibinfo{person}{{\'A}lvaro Brand{\'o}n},
  \bibinfo{person}{Marc Sol{\'e}}, \bibinfo{person}{Alberto Hu{\'e}lamo},
  \bibinfo{person}{David Solans}, \bibinfo{person}{Mar{\'\i}a~S P{\'e}rez},
  {and} \bibinfo{person}{Victor Munt{\'e}s-Mulero}.}
  \bibinfo{year}{2020}\natexlab{}.
\newblock \showarticletitle{Graph-based root cause analysis for
  service-oriented and microservice architectures}.
\newblock \bibinfo{journal}{\emph{Journal of Systems and Software}}
  \bibinfo{volume}{159} (\bibinfo{year}{2020}), \bibinfo{pages}{110432}.
\newblock


\bibitem[\protect\citeauthoryear{Chambers and Thornton}{Chambers and
  Thornton}{2012}]%
        {chambers2012discrete}
\bibfield{author}{\bibinfo{person}{Marcus~J Chambers} {and}
  \bibinfo{person}{Michael~A Thornton}.} \bibinfo{year}{2012}\natexlab{}.
\newblock \showarticletitle{Discrete time representation of continuous time
  ARMA processes}.
\newblock \bibinfo{journal}{\emph{Econometric Theory}} (\bibinfo{year}{2012}),
  \bibinfo{pages}{219--238}.
\newblock


\bibitem[\protect\citeauthoryear{Chen, Kang, Li, Zhang, Zhang, Xu, Zhou, Yang,
  Sun, Xu, et~al\mbox{.}}{Chen et~al\mbox{.}}{2020}]%
        {chen2020towards}
\bibfield{author}{\bibinfo{person}{Zhuangbin Chen}, \bibinfo{person}{Yu Kang},
  \bibinfo{person}{Liqun Li}, \bibinfo{person}{Xu Zhang},
  \bibinfo{person}{Hongyu Zhang}, \bibinfo{person}{Hui Xu},
  \bibinfo{person}{Yangfan Zhou}, \bibinfo{person}{Li Yang},
  \bibinfo{person}{Jeffrey Sun}, \bibinfo{person}{Zhangwei Xu},
  {et~al\mbox{.}}} \bibinfo{year}{2020}\natexlab{}.
\newblock \showarticletitle{Towards intelligent incident management: why we
  need it and how we make it}. In \bibinfo{booktitle}{\emph{Proceedings of the
  28th ACM Joint Meeting on European Software Engineering Conference and
  Symposium on the Foundations of Software Engineering}}.
  \bibinfo{pages}{1487--1497}.
\newblock


\bibitem[\protect\citeauthoryear{Dang, Lin, and Huang}{Dang
  et~al\mbox{.}}{2019}]%
        {dang2019aiops}
\bibfield{author}{\bibinfo{person}{Yingnong Dang}, \bibinfo{person}{Qingwei
  Lin}, {and} \bibinfo{person}{Peng Huang}.} \bibinfo{year}{2019}\natexlab{}.
\newblock \showarticletitle{AIOps: real-world challenges and research
  innovations}. In \bibinfo{booktitle}{\emph{2019 IEEE/ACM 41st International
  Conference on Software Engineering: Companion Proceedings (ICSE-Companion)}}.
  IEEE, \bibinfo{pages}{4--5}.
\newblock


\bibitem[\protect\citeauthoryear{Farshchi, Schneider, Weber, and
  Grundy}{Farshchi et~al\mbox{.}}{2015}]%
        {farshchi2015experience}
\bibfield{author}{\bibinfo{person}{Mostafa Farshchi}, \bibinfo{person}{Jean-Guy
  Schneider}, \bibinfo{person}{Ingo Weber}, {and} \bibinfo{person}{John
  Grundy}.} \bibinfo{year}{2015}\natexlab{}.
\newblock \showarticletitle{Experience report: Anomaly detection of cloud
  application operations using log and cloud metric correlation analysis}. In
  \bibinfo{booktitle}{\emph{2015 IEEE 26th international symposium on software
  reliability engineering (ISSRE)}}. IEEE, \bibinfo{pages}{24--34}.
\newblock


\bibitem[\protect\citeauthoryear{Flink}{Flink}{2011}]%
        {flink}
\bibfield{author}{\bibinfo{person}{Apache Flink}.}
  \bibinfo{year}{2011}\natexlab{}.
\newblock \bibinfo{title}{[Online]}.
\newblock \bibinfo{howpublished}{\url{https://flink.apache.org/}}.
\newblock


\bibitem[\protect\citeauthoryear{Frey and Dueck}{Frey and Dueck}{2007}]%
        {frey2007clustering}
\bibfield{author}{\bibinfo{person}{Brendan~J Frey} {and}
  \bibinfo{person}{Delbert Dueck}.} \bibinfo{year}{2007}\natexlab{}.
\newblock \showarticletitle{Clustering by passing messages between data
  points}.
\newblock \bibinfo{journal}{\emph{science}} \bibinfo{volume}{315},
  \bibinfo{number}{5814} (\bibinfo{year}{2007}), \bibinfo{pages}{972--976}.
\newblock


\bibitem[\protect\citeauthoryear{Gama, {\v{Z}}liobait{\.e}, Bifet, Pechenizkiy,
  and Bouchachia}{Gama et~al\mbox{.}}{2014}]%
        {gama2014survey}
\bibfield{author}{\bibinfo{person}{Jo{\~a}o Gama}, \bibinfo{person}{Indr{\.e}
  {\v{Z}}liobait{\.e}}, \bibinfo{person}{Albert Bifet}, \bibinfo{person}{Mykola
  Pechenizkiy}, {and} \bibinfo{person}{Abdelhamid Bouchachia}.}
  \bibinfo{year}{2014}\natexlab{}.
\newblock \showarticletitle{A survey on concept drift adaptation}.
\newblock \bibinfo{journal}{\emph{ACM computing surveys (CSUR)}}
  \bibinfo{volume}{46}, \bibinfo{number}{4} (\bibinfo{year}{2014}),
  \bibinfo{pages}{1--37}.
\newblock


\bibitem[\protect\citeauthoryear{Han, Lee, Shen, He, Liu, and Huang}{Han
  et~al\mbox{.}}{2020}]%
        {han2020toward}
\bibfield{author}{\bibinfo{person}{Shujie Han}, \bibinfo{person}{Patrick~PC
  Lee}, \bibinfo{person}{Zhirong Shen}, \bibinfo{person}{Cheng He},
  \bibinfo{person}{Yi Liu}, {and} \bibinfo{person}{Tao Huang}.}
  \bibinfo{year}{2020}\natexlab{}.
\newblock \showarticletitle{Toward adaptive disk failure prediction via stream
  mining}. In \bibinfo{booktitle}{\emph{Proceedings of IEEE ICDCS}}.
\newblock


\bibitem[\protect\citeauthoryear{He, Lin, Lou, Zhang, Lyu, and Zhang}{He
  et~al\mbox{.}}{2018}]%
        {he2018identifying}
\bibfield{author}{\bibinfo{person}{Shilin He}, \bibinfo{person}{Qingwei Lin},
  \bibinfo{person}{Jian-Guang Lou}, \bibinfo{person}{Hongyu Zhang},
  \bibinfo{person}{Michael~R Lyu}, {and} \bibinfo{person}{Dongmei Zhang}.}
  \bibinfo{year}{2018}\natexlab{}.
\newblock \showarticletitle{Identifying impactful service system problems via
  log analysis}. In \bibinfo{booktitle}{\emph{Proceedings of the 2018 26th ACM
  Joint Meeting on European Software Engineering Conference and Symposium on
  the Foundations of Software Engineering}}. \bibinfo{pages}{60--70}.
\newblock


\bibitem[\protect\citeauthoryear{Huang, Guo, Zhou, Lorch, Dang, Chintalapati,
  and Yao}{Huang et~al\mbox{.}}{2017}]%
        {huang2017gray}
\bibfield{author}{\bibinfo{person}{Peng Huang}, \bibinfo{person}{Chuanxiong
  Guo}, \bibinfo{person}{Lidong Zhou}, \bibinfo{person}{Jacob~R Lorch},
  \bibinfo{person}{Yingnong Dang}, \bibinfo{person}{Murali Chintalapati}, {and}
  \bibinfo{person}{Randolph Yao}.} \bibinfo{year}{2017}\natexlab{}.
\newblock \showarticletitle{Gray failure: The achilles' heel of cloud-scale
  systems}. In \bibinfo{booktitle}{\emph{Proceedings of the 16th Workshop on
  Hot Topics in Operating Systems}}. \bibinfo{pages}{150--155}.
\newblock


\bibitem[\protect\citeauthoryear{Hundman, Constantinou, Laporte, Colwell, and
  Soderstrom}{Hundman et~al\mbox{.}}{2018}]%
        {hundman2018detecting}
\bibfield{author}{\bibinfo{person}{Kyle Hundman}, \bibinfo{person}{Valentino
  Constantinou}, \bibinfo{person}{Christopher Laporte}, \bibinfo{person}{Ian
  Colwell}, {and} \bibinfo{person}{Tom Soderstrom}.}
  \bibinfo{year}{2018}\natexlab{}.
\newblock \showarticletitle{Detecting spacecraft anomalies using lstms and
  nonparametric dynamic thresholding}. In \bibinfo{booktitle}{\emph{Proceedings
  of the 24th ACM SIGKDD international conference on knowledge discovery \&
  data mining}}. \bibinfo{pages}{387--395}.
\newblock


\bibitem[\protect\citeauthoryear{Kafka}{Kafka}{2011}]%
        {kafka}
\bibfield{author}{\bibinfo{person}{Apache Kafka}.}
  \bibinfo{year}{2011}\natexlab{}.
\newblock \bibinfo{title}{[Online]}.
\newblock \bibinfo{howpublished}{\url{https://kafka.apache.org/}}.
\newblock


\bibitem[\protect\citeauthoryear{Lin, Hsieh, Dang, Zhang, Sui, Xu, Lou, Li, Wu,
  Yao, et~al\mbox{.}}{Lin et~al\mbox{.}}{2018}]%
        {lin2018predicting}
\bibfield{author}{\bibinfo{person}{Qingwei Lin}, \bibinfo{person}{Ken Hsieh},
  \bibinfo{person}{Yingnong Dang}, \bibinfo{person}{Hongyu Zhang},
  \bibinfo{person}{Kaixin Sui}, \bibinfo{person}{Yong Xu},
  \bibinfo{person}{Jian-Guang Lou}, \bibinfo{person}{Chenggang Li},
  \bibinfo{person}{Youjiang Wu}, \bibinfo{person}{Randolph Yao},
  {et~al\mbox{.}}} \bibinfo{year}{2018}\natexlab{}.
\newblock \showarticletitle{Predicting node failure in cloud service systems}.
  In \bibinfo{booktitle}{\emph{Proceedings of the 2018 26th ACM Joint Meeting
  on European Software Engineering Conference and Symposium on the Foundations
  of Software Engineering}}. \bibinfo{pages}{480--490}.
\newblock


\bibitem[\protect\citeauthoryear{Lin, Zhang, Lou, Zhang, and Chen}{Lin
  et~al\mbox{.}}{2016}]%
        {lin2016log}
\bibfield{author}{\bibinfo{person}{Qingwei Lin}, \bibinfo{person}{Hongyu
  Zhang}, \bibinfo{person}{Jian-Guang Lou}, \bibinfo{person}{Yu Zhang}, {and}
  \bibinfo{person}{Xuewei Chen}.} \bibinfo{year}{2016}\natexlab{}.
\newblock \showarticletitle{Log clustering based problem identification for
  online service systems}. In \bibinfo{booktitle}{\emph{2016 IEEE/ACM 38th
  International Conference on Software Engineering Companion (ICSE-C)}}. IEEE,
  \bibinfo{pages}{102--111}.
\newblock


\bibitem[\protect\citeauthoryear{Liu, Ting, and Zhou}{Liu
  et~al\mbox{.}}{2008}]%
        {liu2008isolation}
\bibfield{author}{\bibinfo{person}{Fei~Tony Liu}, \bibinfo{person}{Kai~Ming
  Ting}, {and} \bibinfo{person}{Zhi-Hua Zhou}.}
  \bibinfo{year}{2008}\natexlab{}.
\newblock \showarticletitle{Isolation forest}. In
  \bibinfo{booktitle}{\emph{2008 eighth ieee international conference on data
  mining}}. IEEE, \bibinfo{pages}{413--422}.
\newblock


\bibitem[\protect\citeauthoryear{Lou, Huang, and Smith}{Lou
  et~al\mbox{.}}{2020}]%
        {lou2020understanding}
\bibfield{author}{\bibinfo{person}{Chang Lou}, \bibinfo{person}{Peng Huang},
  {and} \bibinfo{person}{Scott Smith}.} \bibinfo{year}{2020}\natexlab{}.
\newblock \showarticletitle{Understanding, detecting and localizing partial
  failures in large system software}. In \bibinfo{booktitle}{\emph{17th USENIX
  Symposium on Networked Systems Design and Implementation (NSDI 20)}}.
  \bibinfo{pages}{559--574}.
\newblock


\bibitem[\protect\citeauthoryear{Lou, Lin, Ding, Fu, Zhang, and Xie}{Lou
  et~al\mbox{.}}{2013}]%
        {lou2013software}
\bibfield{author}{\bibinfo{person}{Jian-Guang Lou}, \bibinfo{person}{Qingwei
  Lin}, \bibinfo{person}{Rui Ding}, \bibinfo{person}{Qiang Fu},
  \bibinfo{person}{Dongmei Zhang}, {and} \bibinfo{person}{Tao Xie}.}
  \bibinfo{year}{2013}\natexlab{}.
\newblock \showarticletitle{Software analytics for incident management of
  online services: An experience report}. In \bibinfo{booktitle}{\emph{2013
  28th IEEE/ACM International Conference on Automated Software Engineering
  (ASE)}}. IEEE, \bibinfo{pages}{475--485}.
\newblock


\bibitem[\protect\citeauthoryear{Ma, Yin, Zhang, Wang, Zheng, Jiang, Hu, Luo,
  Li, Qiu, et~al\mbox{.}}{Ma et~al\mbox{.}}{2020}]%
        {ma2020diagnosing}
\bibfield{author}{\bibinfo{person}{Minghua Ma}, \bibinfo{person}{Zheng Yin},
  \bibinfo{person}{Shenglin Zhang}, \bibinfo{person}{Sheng Wang},
  \bibinfo{person}{Christopher Zheng}, \bibinfo{person}{Xinhao Jiang},
  \bibinfo{person}{Hanwen Hu}, \bibinfo{person}{Cheng Luo},
  \bibinfo{person}{Yilin Li}, \bibinfo{person}{Nengjun Qiu}, {et~al\mbox{.}}}
  \bibinfo{year}{2020}\natexlab{}.
\newblock \showarticletitle{Diagnosing root causes of intermittent slow queries
  in cloud databases}.
\newblock \bibinfo{journal}{\emph{Proceedings of the VLDB Endowment}}
  \bibinfo{volume}{13}, \bibinfo{number}{10} (\bibinfo{year}{2020}),
  \bibinfo{pages}{1176--1189}.
\newblock


\bibitem[\protect\citeauthoryear{Mercer, Alaee, Abdoli, Singh, Murillo, and
  Keogh}{Mercer et~al\mbox{.}}{2021}]%
        {mercer2021matrix}
\bibfield{author}{\bibinfo{person}{Ryan Mercer}, \bibinfo{person}{Sara Alaee},
  \bibinfo{person}{Alireza Abdoli}, \bibinfo{person}{Shailendra Singh},
  \bibinfo{person}{Amy Murillo}, {and} \bibinfo{person}{Eamonn Keogh}.}
  \bibinfo{year}{2021}\natexlab{}.
\newblock \showarticletitle{Matrix Profile XXIII: Contrast Profile: A Novel
  Time Series Primitive that Allows Real World Classification}. In
  \bibinfo{booktitle}{\emph{The IEEE International Conference on Data Mining}}.
\newblock


\bibitem[\protect\citeauthoryear{Pang, Ting, and Albrecht}{Pang
  et~al\mbox{.}}{2015}]%
        {pang2015lesinn}
\bibfield{author}{\bibinfo{person}{Guansong Pang}, \bibinfo{person}{Kai~Ming
  Ting}, {and} \bibinfo{person}{David~W. Albrecht}.}
  \bibinfo{year}{2015}\natexlab{}.
\newblock \showarticletitle{LeSiNN: Detecting Anomalies by Identifying Least
  Similar Nearest Neighbours}. In \bibinfo{booktitle}{\emph{{IEEE}
  International Conference on Data Mining Workshop, {ICDMW} 2015, Atlantic
  City, NJ, USA, November 14-17, 2015}}. \bibinfo{publisher}{{IEEE} Computer
  Society}, \bibinfo{pages}{623--630}.
\newblock


\bibitem[\protect\citeauthoryear{Park, Hoshi, and Kemp}{Park
  et~al\mbox{.}}{2018}]%
        {park2018multimodal}
\bibfield{author}{\bibinfo{person}{Daehyung Park}, \bibinfo{person}{Yuuna
  Hoshi}, {and} \bibinfo{person}{Charles~C Kemp}.}
  \bibinfo{year}{2018}\natexlab{}.
\newblock \showarticletitle{A multimodal anomaly detector for robot-assisted
  feeding using an lstm-based variational autoencoder}.
\newblock \bibinfo{journal}{\emph{IEEE Robotics and Automation Letters}}
  \bibinfo{volume}{3}, \bibinfo{number}{3} (\bibinfo{year}{2018}),
  \bibinfo{pages}{1544--1551}.
\newblock


\bibitem[\protect\citeauthoryear{Pevn{\`y}}{Pevn{\`y}}{2016}]%
        {pevny2016loda}
\bibfield{author}{\bibinfo{person}{Tom{\'a}{\v{s}} Pevn{\`y}}.}
  \bibinfo{year}{2016}\natexlab{}.
\newblock \showarticletitle{Loda: Lightweight on-line detector of anomalies}.
\newblock \bibinfo{journal}{\emph{Machine Learning}} \bibinfo{volume}{102},
  \bibinfo{number}{2} (\bibinfo{year}{2016}), \bibinfo{pages}{275--304}.
\newblock


\bibitem[\protect\citeauthoryear{Podgurski, Leon, Francis, Masri, Minch, Sun,
  and Wang}{Podgurski et~al\mbox{.}}{2003}]%
        {podgurski2003automated}
\bibfield{author}{\bibinfo{person}{Andy Podgurski}, \bibinfo{person}{David
  Leon}, \bibinfo{person}{Patrick Francis}, \bibinfo{person}{Wes Masri},
  \bibinfo{person}{Melinda Minch}, \bibinfo{person}{Jiayang Sun}, {and}
  \bibinfo{person}{Bin Wang}.} \bibinfo{year}{2003}\natexlab{}.
\newblock \showarticletitle{Automated support for classifying software failure
  reports}. In \bibinfo{booktitle}{\emph{25th International Conference on
  Software Engineering, 2003. Proceedings.}} IEEE, \bibinfo{pages}{465--475}.
\newblock


\bibitem[\protect\citeauthoryear{Rakthanmanon and Keogh}{Rakthanmanon and
  Keogh}{2013}]%
        {rakthanmanon2013fast}
\bibfield{author}{\bibinfo{person}{Thanawin Rakthanmanon} {and}
  \bibinfo{person}{Eamonn Keogh}.} \bibinfo{year}{2013}\natexlab{}.
\newblock \showarticletitle{Fast shapelets: A scalable algorithm for
  discovering time series shapelets}. In \bibinfo{booktitle}{\emph{proceedings
  of the 2013 SIAM International Conference on Data Mining}}. SIAM,
  \bibinfo{pages}{668--676}.
\newblock


\bibitem[\protect\citeauthoryear{Ren, Xu, Wang, Yi, Huang, Kou, Xing, Yang,
  Tong, and Zhang}{Ren et~al\mbox{.}}{2019}]%
        {ren2019time}
\bibfield{author}{\bibinfo{person}{Hansheng Ren}, \bibinfo{person}{Bixiong Xu},
  \bibinfo{person}{Yujing Wang}, \bibinfo{person}{Chao Yi},
  \bibinfo{person}{Congrui Huang}, \bibinfo{person}{Xiaoyu Kou},
  \bibinfo{person}{Tony Xing}, \bibinfo{person}{Mao Yang}, \bibinfo{person}{Jie
  Tong}, {and} \bibinfo{person}{Qi Zhang}.} \bibinfo{year}{2019}\natexlab{}.
\newblock \showarticletitle{Time-series anomaly detection service at
  microsoft}. In \bibinfo{booktitle}{\emph{Proceedings of the 25th ACM SIGKDD
  International Conference on Knowledge Discovery \& Data Mining}}.
  \bibinfo{pages}{3009--3017}.
\newblock


\bibitem[\protect\citeauthoryear{Research}{Research}{2015}]%
        {yahoo}
\bibfield{author}{\bibinfo{person}{Yahoo! Research}.}
  \bibinfo{year}{2015}\natexlab{}.
\newblock \bibinfo{title}{A Benchmark Dataset for Time Series Anomaly
  Detection}.
\newblock
\newblock
\urldef\tempurl%
\url{https://yahooresearch.tumblr.com/post/114590420346/a-benchmark-dataset-for-time-series-anomaly}
\showURL{%
Retrieved August, 2021 from \tempurl}


\bibitem[\protect\citeauthoryear{Siffer, Fouque, Termier, and
  Largou{\"{e}}t}{Siffer et~al\mbox{.}}{2017}]%
        {siffer2017anomaly}
\bibfield{author}{\bibinfo{person}{Alban Siffer},
  \bibinfo{person}{Pierre{-}Alain Fouque}, \bibinfo{person}{Alexandre Termier},
  {and} \bibinfo{person}{Christine Largou{\"{e}}t}.}
  \bibinfo{year}{2017}\natexlab{}.
\newblock \showarticletitle{Anomaly Detection in Streams with Extreme Value
  Theory}. In \bibinfo{booktitle}{\emph{Proceedings of the 23rd {ACM} {SIGKDD}
  International Conference on Knowledge Discovery and Data Mining, Halifax, NS,
  Canada, August 13 - 17, 2017}}. \bibinfo{publisher}{{ACM}},
  \bibinfo{pages}{1067--1075}.
\newblock


\bibitem[\protect\citeauthoryear{Su, Zhao, Niu, Liu, Sun, and Pei}{Su
  et~al\mbox{.}}{2019}]%
        {su2019robust}
\bibfield{author}{\bibinfo{person}{Ya Su}, \bibinfo{person}{Youjian Zhao},
  \bibinfo{person}{Chenhao Niu}, \bibinfo{person}{Rong Liu},
  \bibinfo{person}{Wei Sun}, {and} \bibinfo{person}{Dan Pei}.}
  \bibinfo{year}{2019}\natexlab{}.
\newblock \showarticletitle{Robust anomaly detection for multivariate time
  series through stochastic recurrent neural network}. In
  \bibinfo{booktitle}{\emph{Proceedings of the 25th ACM SIGKDD International
  Conference on Knowledge Discovery \& Data Mining}}.
  \bibinfo{pages}{2828--2837}.
\newblock


\bibitem[\protect\citeauthoryear{Trubiani, Jamshidi, Cito, Shang, Jiang, and
  Borg}{Trubiani et~al\mbox{.}}{2018}]%
        {trubiani2018performance}
\bibfield{author}{\bibinfo{person}{Catia Trubiani}, \bibinfo{person}{Pooyan
  Jamshidi}, \bibinfo{person}{Jurgen Cito}, \bibinfo{person}{Weiyi Shang},
  \bibinfo{person}{Zhen~Ming Jiang}, {and} \bibinfo{person}{Markus Borg}.}
  \bibinfo{year}{2018}\natexlab{}.
\newblock \showarticletitle{Performance issues? Hey DevOps, mind the
  uncertainty}.
\newblock \bibinfo{journal}{\emph{IEEE Software}} \bibinfo{volume}{36},
  \bibinfo{number}{2} (\bibinfo{year}{2018}), \bibinfo{pages}{110--117}.
\newblock


\bibitem[\protect\citeauthoryear{Xu, Chen, Zhao, Li, Bu, Li, Liu, Zhao, Pei,
  Feng, et~al\mbox{.}}{Xu et~al\mbox{.}}{2018}]%
        {xu2018unsupervised}
\bibfield{author}{\bibinfo{person}{Haowen Xu}, \bibinfo{person}{Wenxiao Chen},
  \bibinfo{person}{Nengwen Zhao}, \bibinfo{person}{Zeyan Li},
  \bibinfo{person}{Jiahao Bu}, \bibinfo{person}{Zhihan Li},
  \bibinfo{person}{Ying Liu}, \bibinfo{person}{Youjian Zhao},
  \bibinfo{person}{Dan Pei}, \bibinfo{person}{Yang Feng}, {et~al\mbox{.}}}
  \bibinfo{year}{2018}\natexlab{}.
\newblock \showarticletitle{Unsupervised anomaly detection via variational
  auto-encoder for seasonal kpis in web applications}. In
  \bibinfo{booktitle}{\emph{Proceedings of the 2018 World Wide Web
  Conference}}. \bibinfo{pages}{187--196}.
\newblock


\bibitem[\protect\citeauthoryear{Yankov, Keogh, and Rebbapragada}{Yankov
  et~al\mbox{.}}{2008}]%
        {yankov2008disk}
\bibfield{author}{\bibinfo{person}{Dragomir Yankov}, \bibinfo{person}{Eamonn
  Keogh}, {and} \bibinfo{person}{Umaa Rebbapragada}.}
  \bibinfo{year}{2008}\natexlab{}.
\newblock \showarticletitle{Disk aware discord discovery: Finding unusual time
  series in terabyte sized datasets}.
\newblock \bibinfo{journal}{\emph{Knowledge and Information Systems}}
  \bibinfo{volume}{17}, \bibinfo{number}{2} (\bibinfo{year}{2008}),
  \bibinfo{pages}{241--262}.
\newblock


\bibitem[\protect\citeauthoryear{Yeh, Zhu, Ulanova, Begum, Ding, Dau, Silva,
  Mueen, and Keogh}{Yeh et~al\mbox{.}}{2016}]%
        {yeh2016matrix}
\bibfield{author}{\bibinfo{person}{Chin-Chia~Michael Yeh}, \bibinfo{person}{Yan
  Zhu}, \bibinfo{person}{Liudmila Ulanova}, \bibinfo{person}{Nurjahan Begum},
  \bibinfo{person}{Yifei Ding}, \bibinfo{person}{Hoang~Anh Dau},
  \bibinfo{person}{Diego~Furtado Silva}, \bibinfo{person}{Abdullah Mueen},
  {and} \bibinfo{person}{Eamonn Keogh}.} \bibinfo{year}{2016}\natexlab{}.
\newblock \showarticletitle{Matrix profile I: all pairs similarity joins for
  time series: a unifying view that includes motifs, discords and shapelets}.
  In \bibinfo{booktitle}{\emph{2016 IEEE 16th international conference on data
  mining (ICDM)}}. IEEE, \bibinfo{pages}{1317--1322}.
\newblock


\bibitem[\protect\citeauthoryear{Zhao, Hassan, Zou, Truong, and Corbin}{Zhao
  et~al\mbox{.}}{2021}]%
        {zhao2021predicting}
\bibfield{author}{\bibinfo{person}{Guoliang Zhao}, \bibinfo{person}{Safwat
  Hassan}, \bibinfo{person}{Ying Zou}, \bibinfo{person}{Derek Truong}, {and}
  \bibinfo{person}{Toby Corbin}.} \bibinfo{year}{2021}\natexlab{}.
\newblock \showarticletitle{Predicting Performance Anomalies in Software
  Systems at Run-time}.
\newblock \bibinfo{journal}{\emph{ACM Transactions on Software Engineering and
  Methodology (TOSEM)}} \bibinfo{volume}{30}, \bibinfo{number}{3}
  (\bibinfo{year}{2021}), \bibinfo{pages}{1--33}.
\newblock


\bibitem[\protect\citeauthoryear{Zhu, Yeh, Zimmerman, Kamgar, and Keogh}{Zhu
  et~al\mbox{.}}{2018}]%
        {zhu2018matrix}
\bibfield{author}{\bibinfo{person}{Yan Zhu}, \bibinfo{person}{Chin-Chia~Michael
  Yeh}, \bibinfo{person}{Zachary Zimmerman}, \bibinfo{person}{Kaveh Kamgar},
  {and} \bibinfo{person}{Eamonn Keogh}.} \bibinfo{year}{2018}\natexlab{}.
\newblock \showarticletitle{Matrix profile XI: SCRIMP++: time series motif
  discovery at interactive speeds}. In \bibinfo{booktitle}{\emph{2018 IEEE
  International Conference on Data Mining (ICDM)}}. IEEE,
  \bibinfo{pages}{837--846}.
\newblock


\bibitem[\protect\citeauthoryear{Zong, Song, Min, Cheng, Lumezanu, Cho, and
  Chen}{Zong et~al\mbox{.}}{2018}]%
        {zong2018deep}
\bibfield{author}{\bibinfo{person}{Bo Zong}, \bibinfo{person}{Qi Song},
  \bibinfo{person}{Martin~Renqiang Min}, \bibinfo{person}{Wei Cheng},
  \bibinfo{person}{Cristian Lumezanu}, \bibinfo{person}{Daeki Cho}, {and}
  \bibinfo{person}{Haifeng Chen}.} \bibinfo{year}{2018}\natexlab{}.
\newblock \showarticletitle{Deep autoencoding gaussian mixture model for
  unsupervised anomaly detection}. In \bibinfo{booktitle}{\emph{International
  Conference on Learning Representations}}.
\newblock


\end{thebibliography}

% \appendix

\end{document}